%% file: LED_arrays_paper.tex
\documentclass[letterpaper,journal]{IEEEtran}

\usepackage[british]{babel}
\usepackage[pdftex]{graphicx}

\usepackage[obeyspaces,hyphens]{url}
\newcommand{\URL}[1]{$\langle$\url{#1}$\rangle$}

\usepackage{setspace} 
\usepackage[caption=false,font=footnotesize]{subfig} 
\usepackage{combelow} 
\usepackage{authblk} 
\usepackage[binary-units]{siunitx} 
\usepackage{array} 
\usepackage{amsmath} 
\usepackage{cite} 
\usepackage[useregional]{datetime2} 
\usepackage{float} 
\usepackage[bookmarks=false]{hyperref} 

\pdfoutput=1 

\newcommand\T{\rule{0pt}{2.6ex}}       
\newcommand\B{\rule[-1.2ex]{0pt}{0pt}} 

\begin{document}

\title{\input{title.tex} \\
\vspace*{20pt}\normalsize\today}

\author{Ireneusz Kubiak\IEEEauthorrefmark{1}
    and Joe Loughry\IEEEauthorrefmark{2} \IEEEmembership{Member,~IEEE}%
\thanks{\IEEEauthorrefmark{1}Military Communication Institute,
Warszawska 22A Str., 05-130 Zegrze, Poland. Email: i.kubiak@wil.waw.pl}
\thanks{\IEEEauthorrefmark{2}University of Denver, \T Ritchie School of
Engineering and Computer Science, 2155 East Wesley Avenue, Denver,
Colorado 80208, USA. Email: joe.loughry@cs.du.edu}}

\maketitle


\begin{abstract}
	\input{abstract.tex}
\end{abstract}


\IEEEpeerreviewmaketitle

\section{Introduction}

\begin{figure}[ht]
    \centering
    \includegraphics[width=\columnwidth]{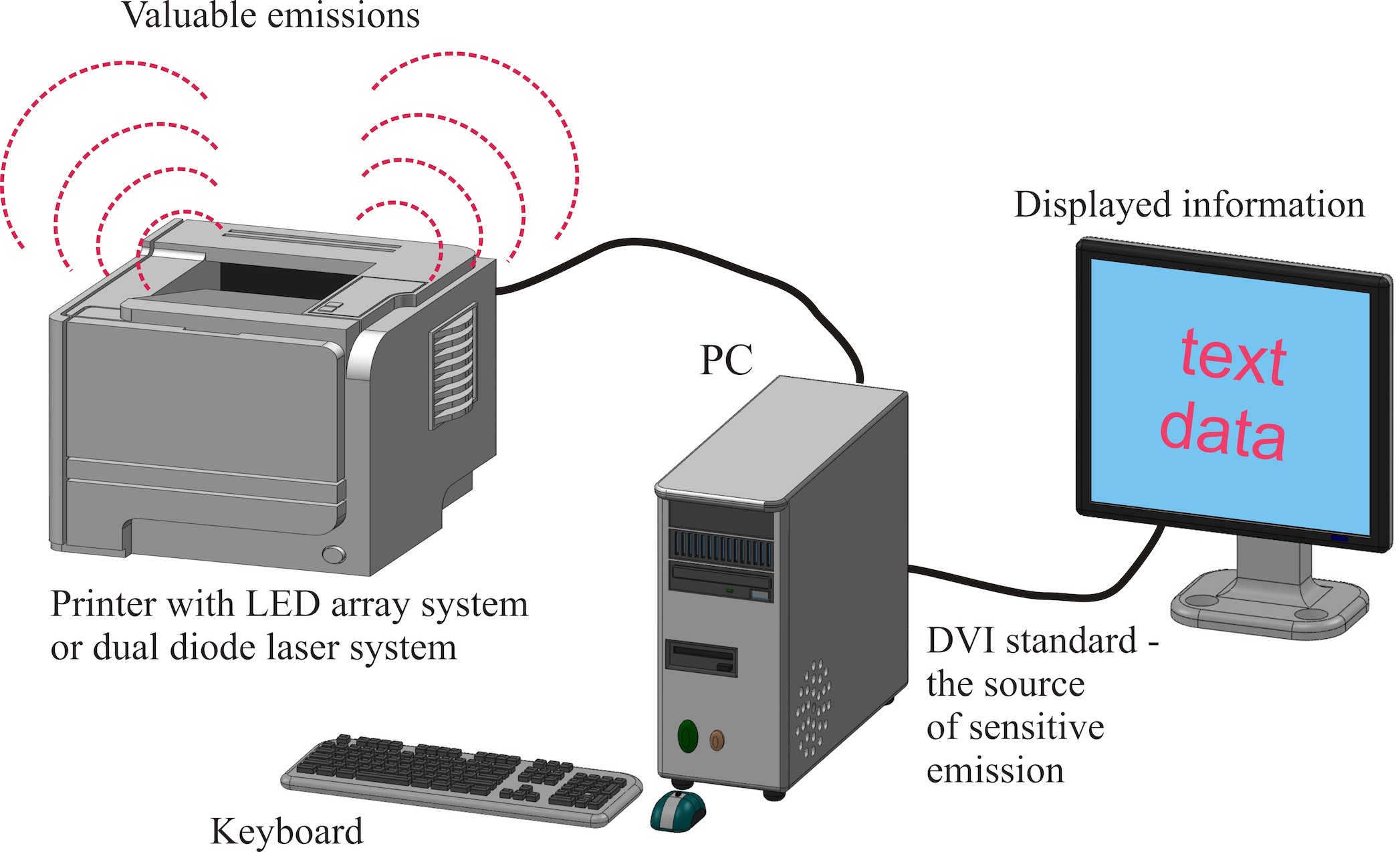}
    \caption{Laser printer as a source of valuable emissions.}
    \label{figure:Figure_01}
\end{figure}

Printers are one of the basic elements of a computer system. They translate
the electronic form of processed data into graphical form during the printing
process. As with every electronic device, printers are sources of
electromagnetic emanations. Besides control signals, which carry no
information ({\it e.g.}, directing the operation of stepper motors or
heaters), there are other signals (useful signals) that are correlated with
the information being processed. Such emissions are called ``sensitive'' or
``valuable'' or ``compromising'' emanations from the point of view of
electromagnetic protection of processed information.\footnote{In fact,
concern over compromising emanations from [electromechanical] printing
devices extends at least as far back as 1956 \cite[[pp.~83--4,
109--111]{Wright1987} or 1917 \cite[Chapter 13]{Kahn1996}.} Processed data may
be information displayed on a computer screen or printed (Figure
\ref{figure:Figure_01}).

Like other devices included in a computer system \cite{Kuhn2002,Kubiak2016a},
the printer can be subject to electromagnetic infiltration, or eavesdropping
\cite{Ketenci2015a,Kubiak2016b}. Therefore, efforts to reduce the level of
susceptibility to electromagnetic eavesdropping are initiated for such
devices. These are unintentionally radiated signals; by `infiltration' we
mean exploitation of naturally occurring intelligence-bearing modulation, not
the introduction of deliberately induced vulnerabilities. It is more like
\textsc{engulf} than \textsc{gunman} \cite{Wright1987,Maneki2007a}.
Organisational and technical solutions\footnote{An example of an
organisational solution might be establishment of a ``control zone'' around
susceptible devices, relying on distance to attenuate signals below levels
that can be received outside the control zone.} are the most often-used
methods for limiting infiltration sensitivity of devices \cite{Kubiak2006a}.
Technical solutions are limited to changes in the
design of devices that typically increase the cost of such devices and
sometimes limit their functionality. Therefore, it is desirable to find
solutions that avoid these drawbacks and at the same time allow ``safe''
processing of classified information \cite{Wasfy2011a,Goel2012a}.

One technical method that is commonly used in the field of electromagnetic
compatibility---both to reduce the amount of electromagnetic interference
emitted from the device and the susceptibility of the device to
electromagnetic disturbance---is the use of differential-mode signals.
In this paper, analysis of useful signals and control signals
\cite{Kubiak2017d} in the operation of LED arrays
used in printers shows that such a design was used by printer $B$ in the
operation of its photoconductor exposure system. Is this sufficient, however,
to foil non-invasive information gathering? Research and results are
presented in this article.

The clear answer is that the solution adopted in the design of the $B$
printer (Figure \ref{figure:Figure_02b}) significantly reduces the
susceptibility of the device to infiltration, in comparison to the $A$
printer (Figure \ref{figure:Figure_02a}). Moreover, the level of
electromagnetic emission of printer $A$ is higher than that of
typical single or dual diode laser printers \cite{Kubiak2014b}.

\section{The Structure of the Emission Source and the Characteristics of
Useful Signals}

\begin{figure*}[!ht]
    \centering
    \subfloat[]{\includegraphics[width=\columnwidth]{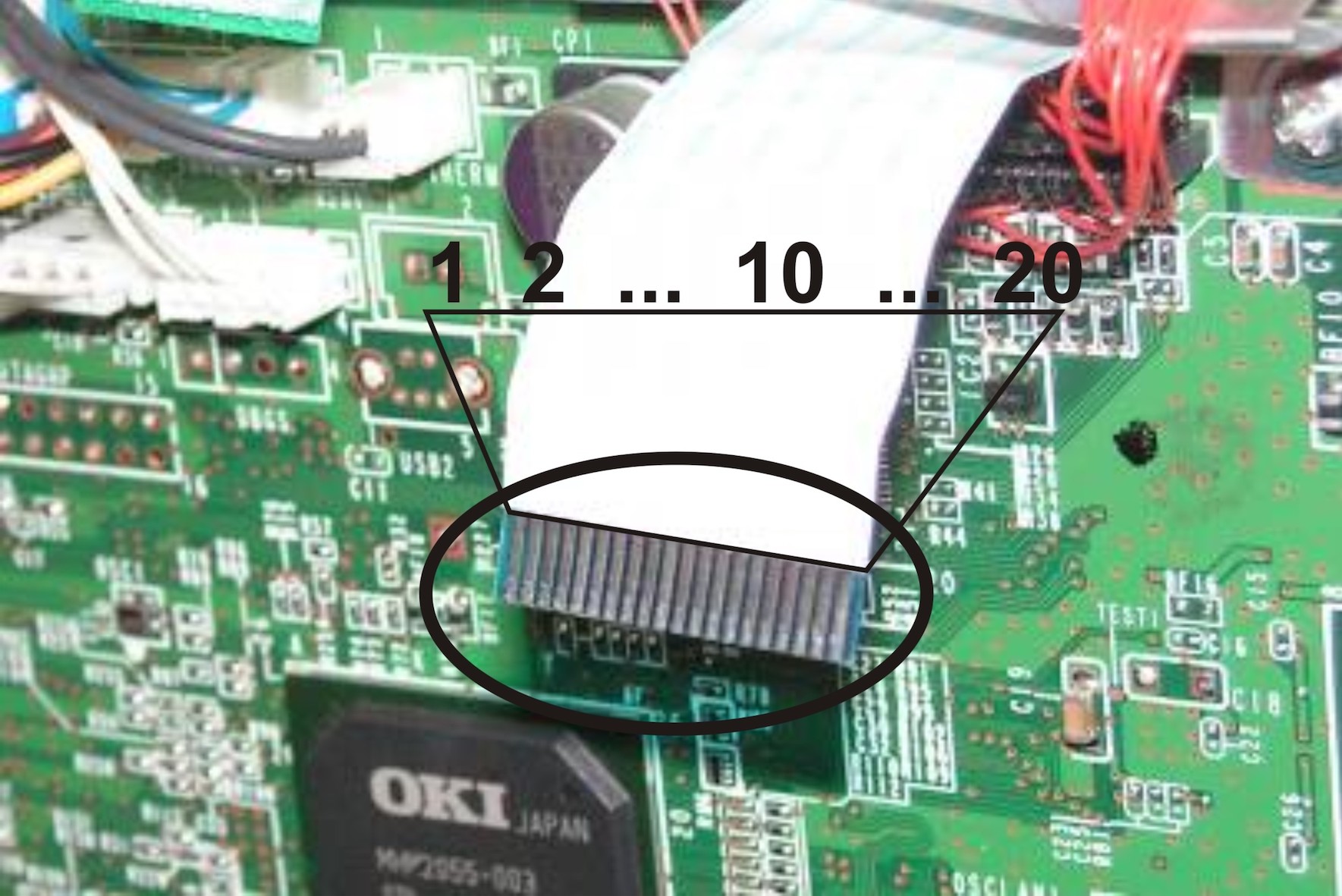}%
    \label{figure:Figure_03a}}
    \hfill
    \subfloat[]{\includegraphics[width=\columnwidth]{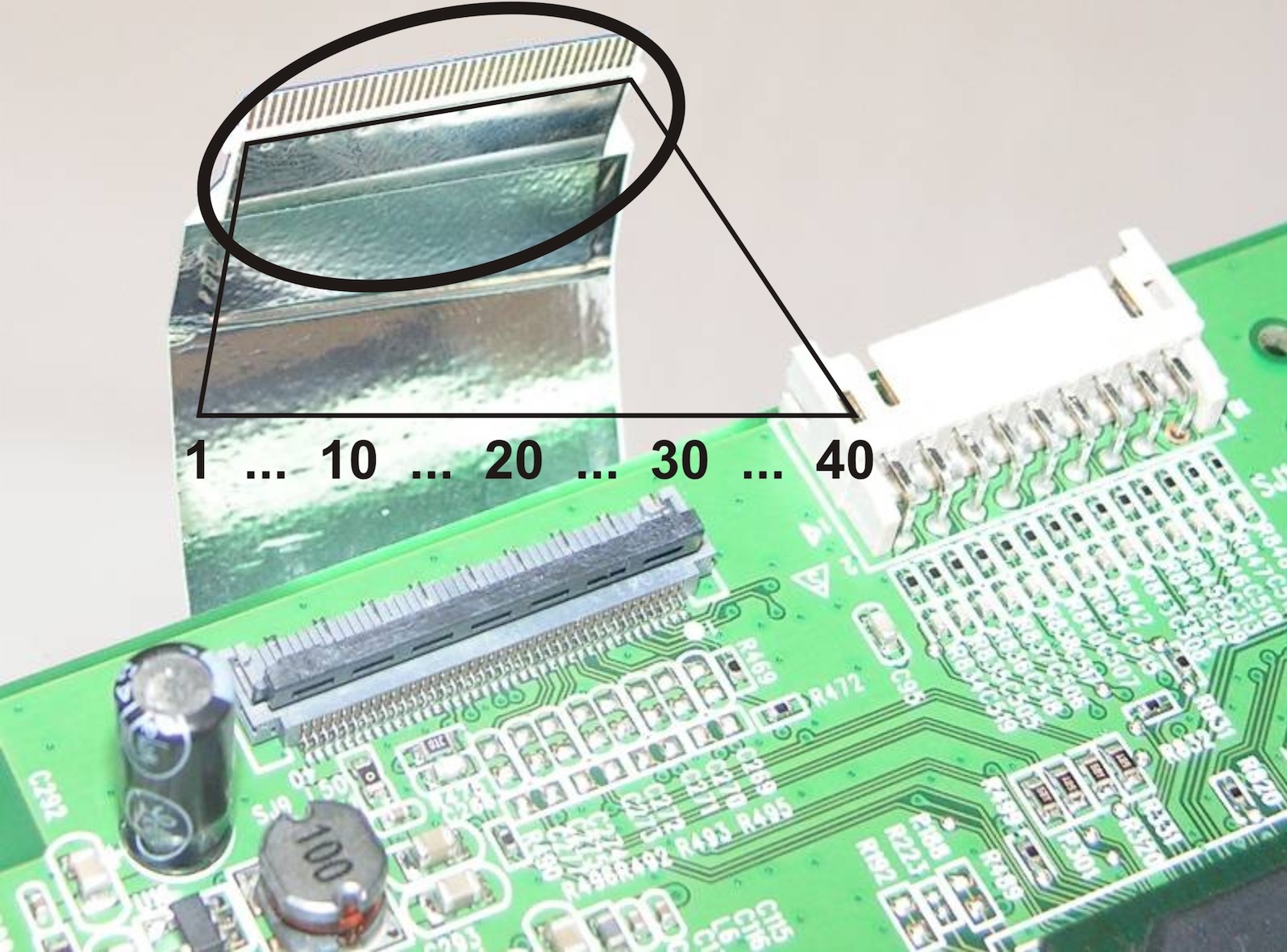}%
    \label{figure:Figure_03b}}
    \caption{Ribbon cable supplying useful signals to the LED array: (a)
        Printer $A$, (b) Printer $B$.}
    \label{figure:Figure_03}
\end{figure*}

The analyses were carried out on two printers using LED array technology.
Different ways of controlling the LED array---chosen by the printer's
designer---affect the number of useful signals (Figure
\ref{figure:Figure_03}) and the structure of those signals. In the case of
printer $A$ we can distinguish four useful signals and six control
signals. The next ten wires are ground wires. Printer $B$ has eight
useful signals (four differential pairs).

\begin{figure*}[ht]
    \centering
    \subfloat[]{\includegraphics[width=\columnwidth]{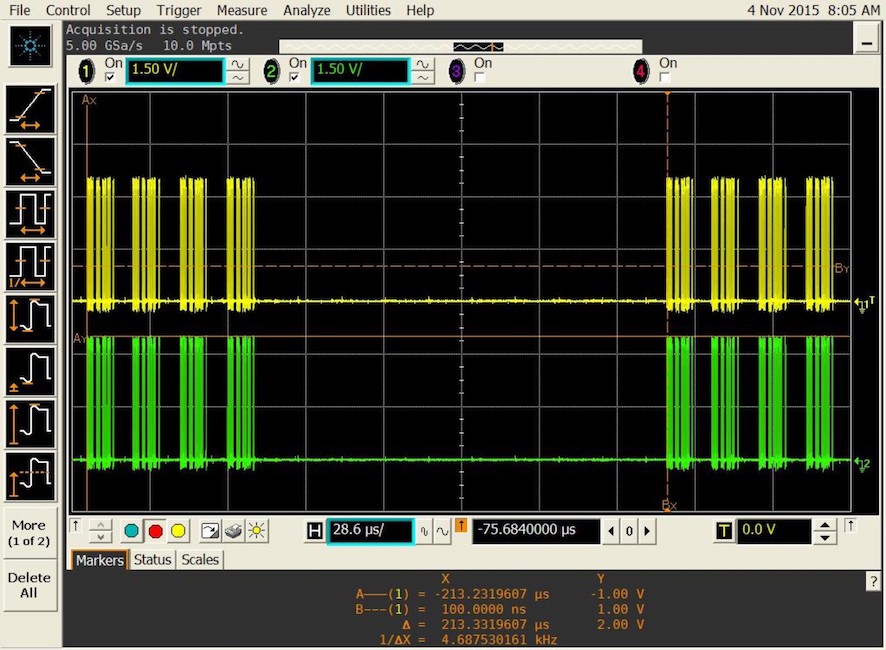}%
    \label{figure:Figure_04a}}
    \hfill
    \subfloat[]{\includegraphics[width=\columnwidth]{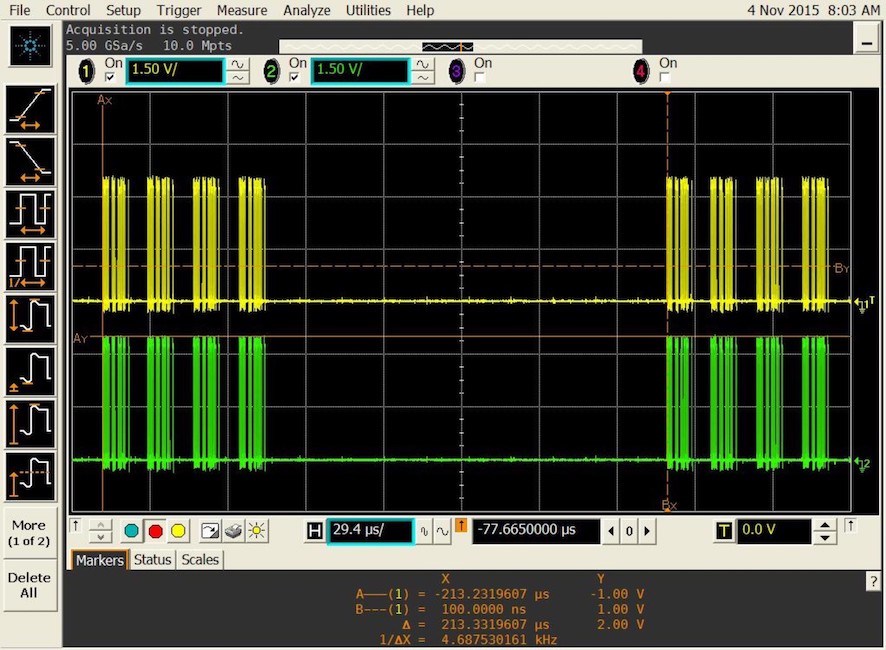}%
    \label{figure:Figure_04b}}
    \caption{Waveforms of useful signals on pins 2 (lower trace) and 5
        (upper trace) of printer $A$ for: a) the 300\,dpi mode and the
        Best option, b) the 300\,dpi mode and the Eco option.}

    \label{figure:Figure_04}
\end{figure*}

\begin{figure*}[ht]
    \centering
    \subfloat[]{\includegraphics[width=\columnwidth]{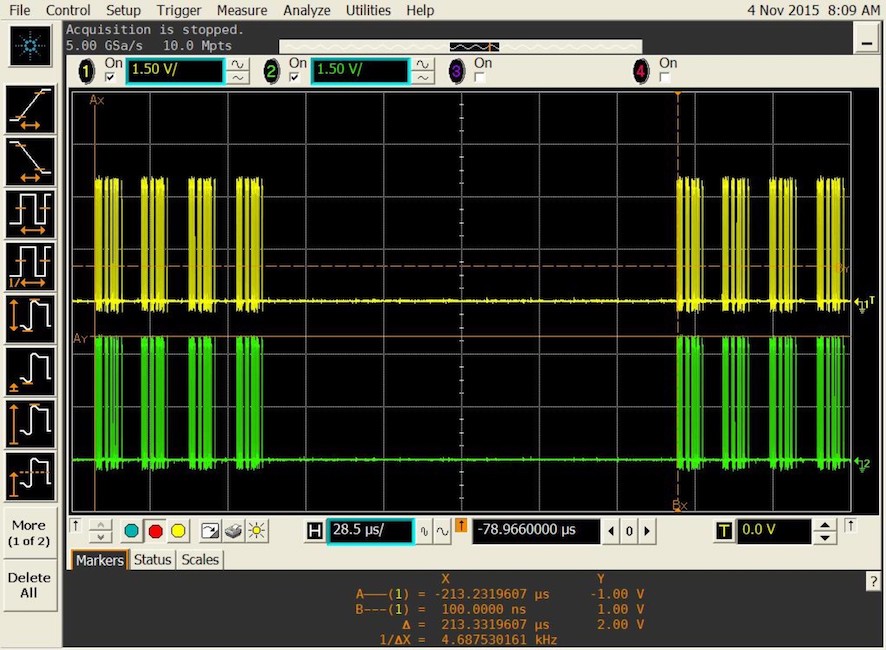}%
    \label{figure:Figure_05a}}
    \hfill
    \subfloat[]{\includegraphics[width=\columnwidth]{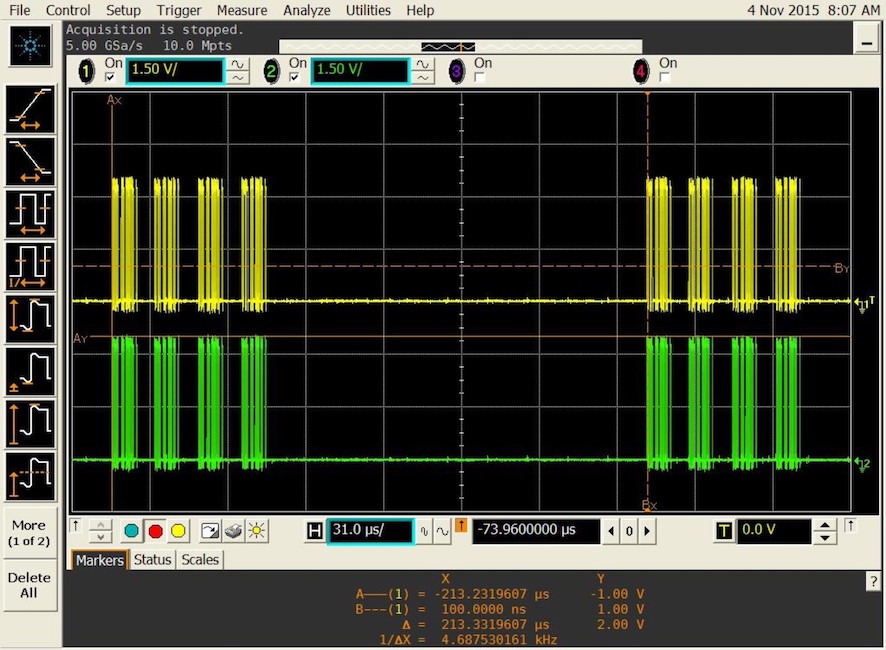}%
    \label{figure:Figure_05b}}
    \caption{Waveforms of useful signals on pins 2 (lower trace) and 3 (upper
        trace) of printer $A$ for: a) the 600\,dpi mode and the Best
        option, b) the 600\,dpi mode and the Eco option.}
    \label{figure:Figure_05}
\end{figure*}

\begin{figure*}[ht]
    \centering
    \subfloat[]{\includegraphics[width=\columnwidth]{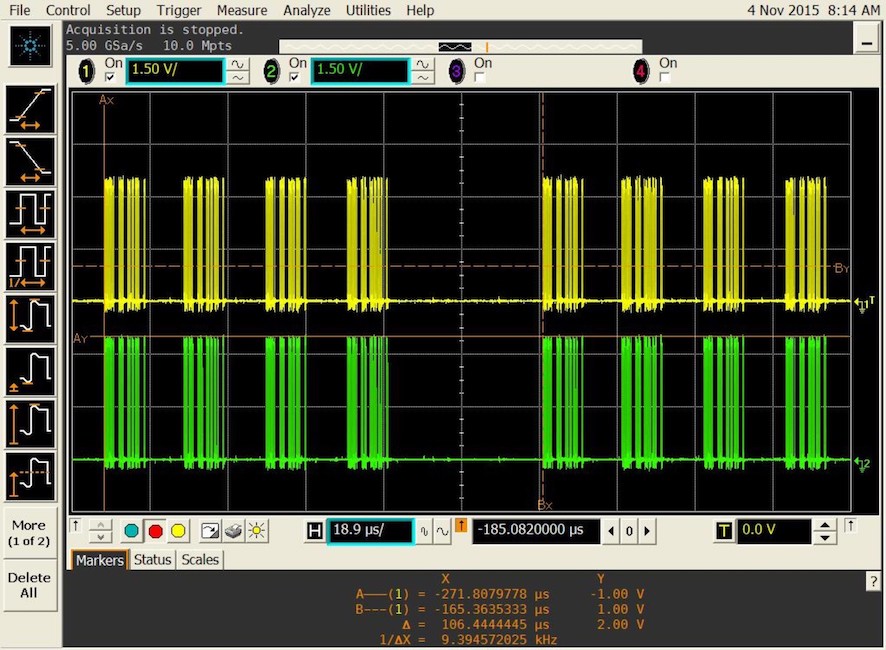}%
    \label{figure:Figure_06a}}
    \hfill
    \subfloat[]{\includegraphics[width=\columnwidth]{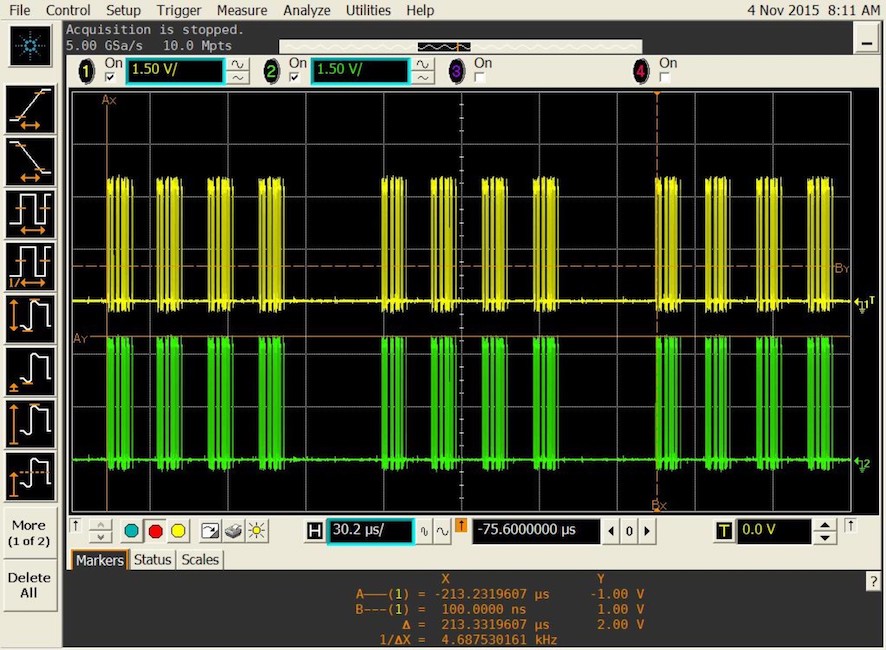}%
    \label{figure:Figure_06b}}
    \caption{Waveforms of useful signals on pins 2 (lower trace) and 3 (upper
        trace) of printer $A$ for: a) the 1200\,dpi mode and the Best
        option, b) the 1200\,dpi mode and the Eco option.}
    \label{figure:Figure_06}
\end{figure*}

The other signals are control wires and ground wires (32 in all). By
probing signal wires while exercising the printer, we were able to learn the
structure of the control signals, how the LED array is controlled, and the
way in which different print quality
options are achieved depending on the operating mode and the ``toner save''
option. Each of the tested printers uses different methods of controlling the
LED array, which can affect the level of electromagnetic emanations. Examples
of waveforms of useful and control signals for printer $A$ are shown
in the oscilloscope traces of Figures
\ref{figure:Figure_04}--\ref{figure:Figure_06}.

The structures of useful signals, based on the example of the signal on pin
2, does not change for the 300\,dpi and 600\,dpi operating modes of the
printer. In the case of the 1200\,dpi mode, the frequency of signal repetition
increases by two. The amplitude is constant at approximately \SI{3.5}{\volt}
(Table~\ref{table:Table_1}).

%
%

\begin{table}[!t]
    \caption{Parameters of useful signals of printer $A$ in relation to
        printing parameters.}
    \label{table:Table_1}
    \centering
    \begin{tabular}{|c|c|c|}
        \hline
        & \multicolumn{2}{c|}{Parameters of Useful Signal}\T\B \\
        \hline
        Operating Mode & Frequency (\si{\kilo\hertz})
                       & Amplitude (\si{\volt})\T\B \\
        \hline
        300\,dpi, Eco\T & $\sim 4.7$ & 3.5 \\
        300\,dpi, Best  & $\sim 4.7$ & 3.5 \\
        600\,dpi, Eco   & $\sim 4.7$ & 3.5 \\
        600\,dpi, Best  & $\sim 4.7$ & 3.5 \\
        1200\,dpi, Eco  & $\sim 9.4$ & 3.5 \\
        1200\,dpi, Best & $\sim 9.4$ & 3.5\B \\
        \hline
    \end{tabular}
\end{table}

The structure of the signal (waveform shape and duty cycle) doesn't change.
This proves that the level of risk of electromagnetic emanations correlated
with the processed (printed) information is not affected by printing quality
(resolution and toner save option), in contrast to the situation found with
single and dual diode laser printers \cite{Kubiak2016b}.

\begin{figure}[ht]
    \centering
    \includegraphics[width=\columnwidth]{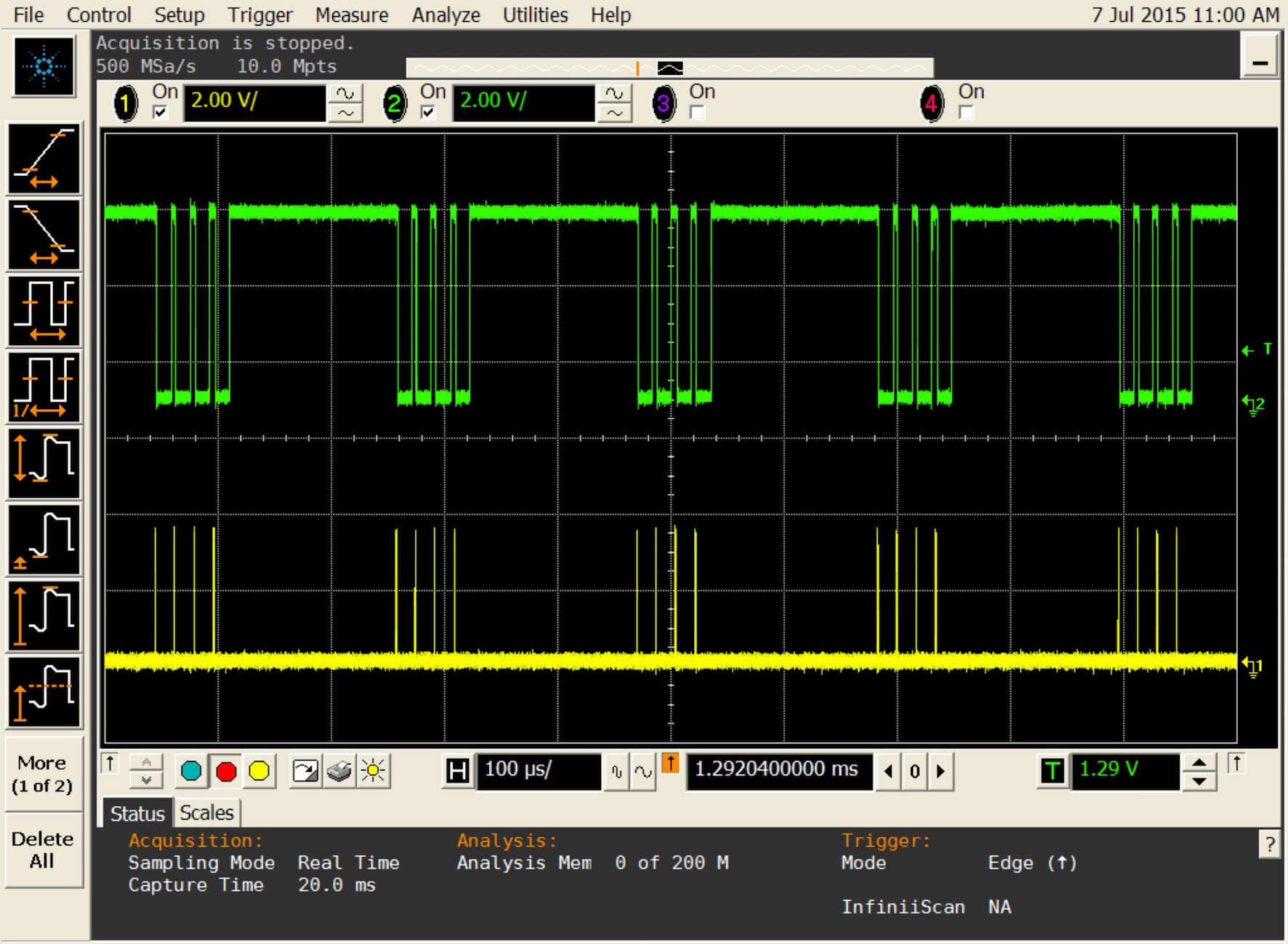}
    \caption{Waveforms of control signals on pins 6 (lower trace) and 9
        (upper trace) of printer $A$ for the 300\,dpi mode and the Best
        option.}
    \label{figure:Figure_07}
\end{figure}

\begin{figure}[ht]
    \centering
    \includegraphics[width=\columnwidth]{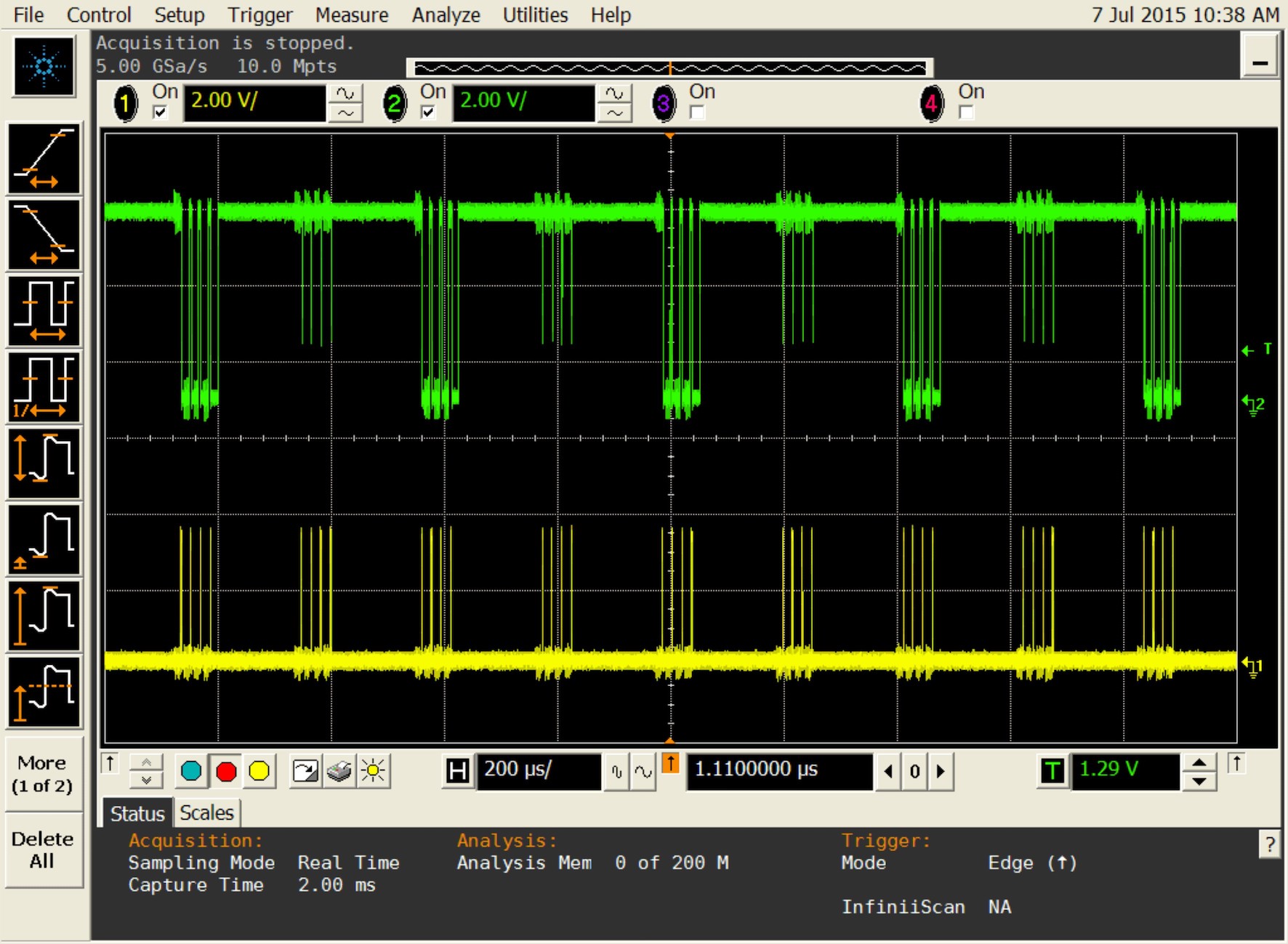}
    \caption{Waveforms of control signals on pins 6 (lower trace) and 9
        (upper trace) of printer $A$ for the 300\,dpi mode and the Eco
        option.}
    \label{figure:Figure_08}
\end{figure}

\begin{figure}[ht]
    \centering
    \includegraphics[width=\columnwidth]{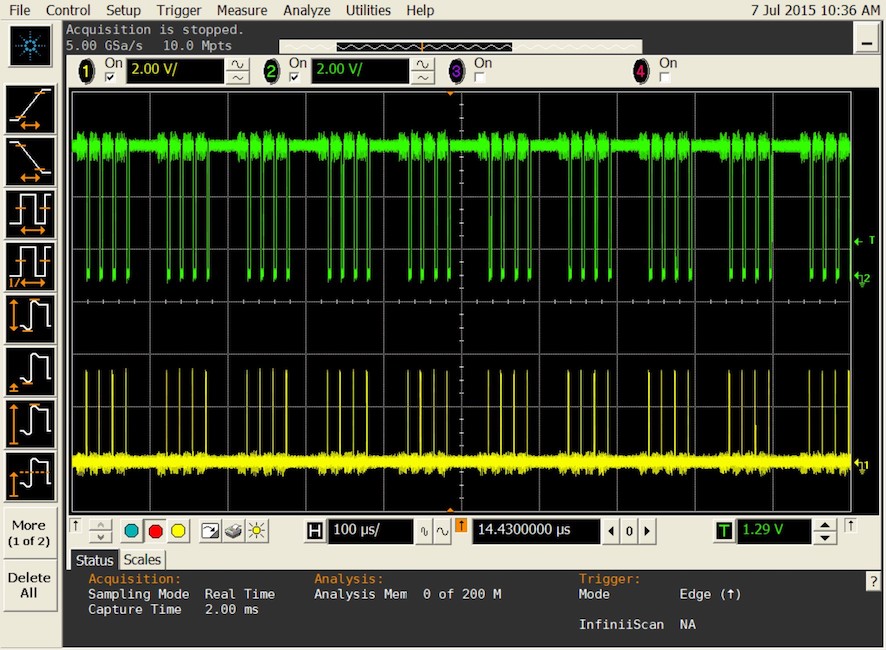}
    \caption{Waveforms of control signals on pins 6 (lower trace) and 9
    (upper trace) of printer $A$ in 1200\,dpi mode with the Best option.}
    \label{figure:Figure_09}
\end{figure}

\begin{figure}[H] 
    \centering
    \subfloat[Printer $A$]{\includegraphics[width=0.45\columnwidth]{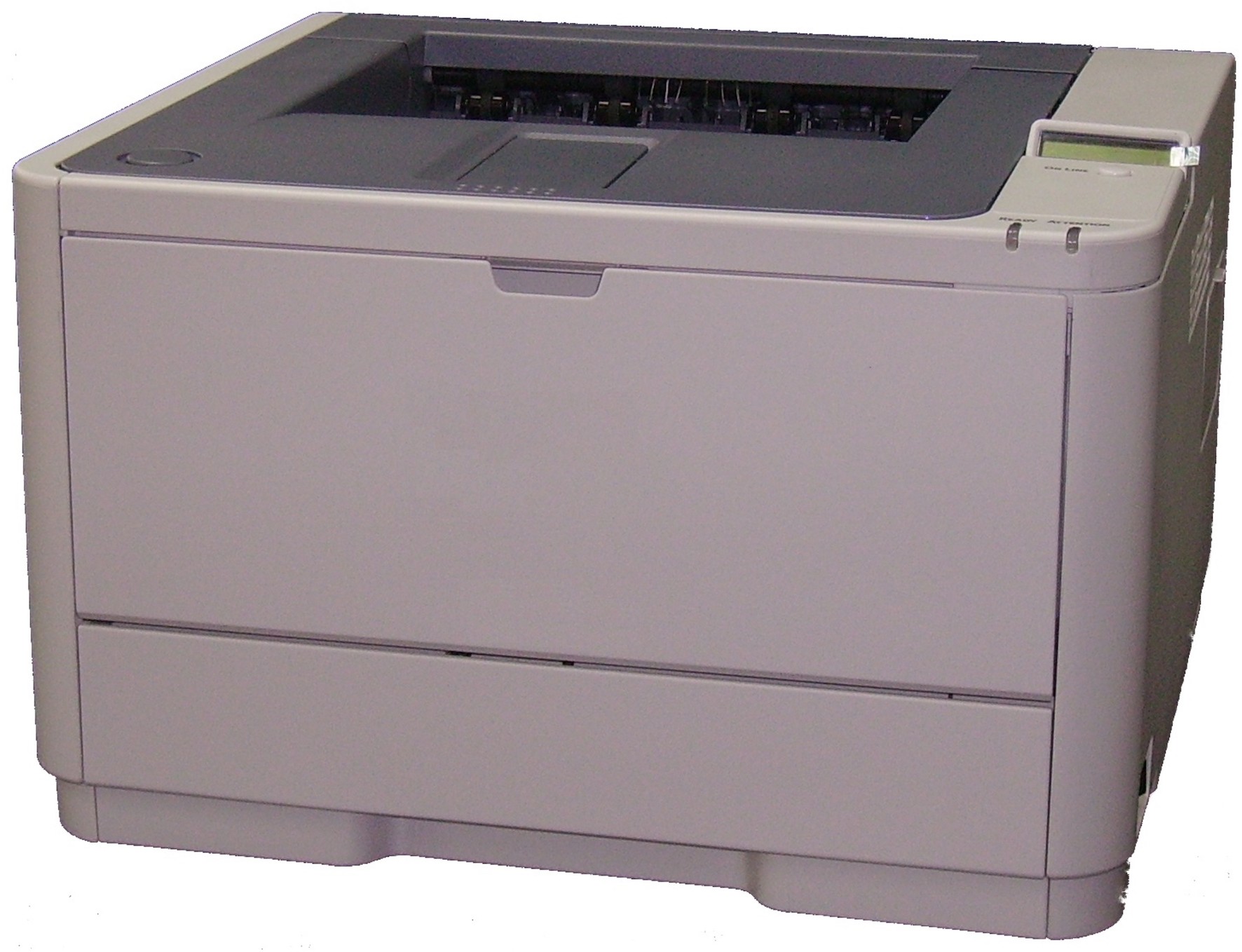}%
    \label{figure:Figure_02a}}
    \hfill
    \subfloat[Printer $B$]{\includegraphics[width=0.45\columnwidth]{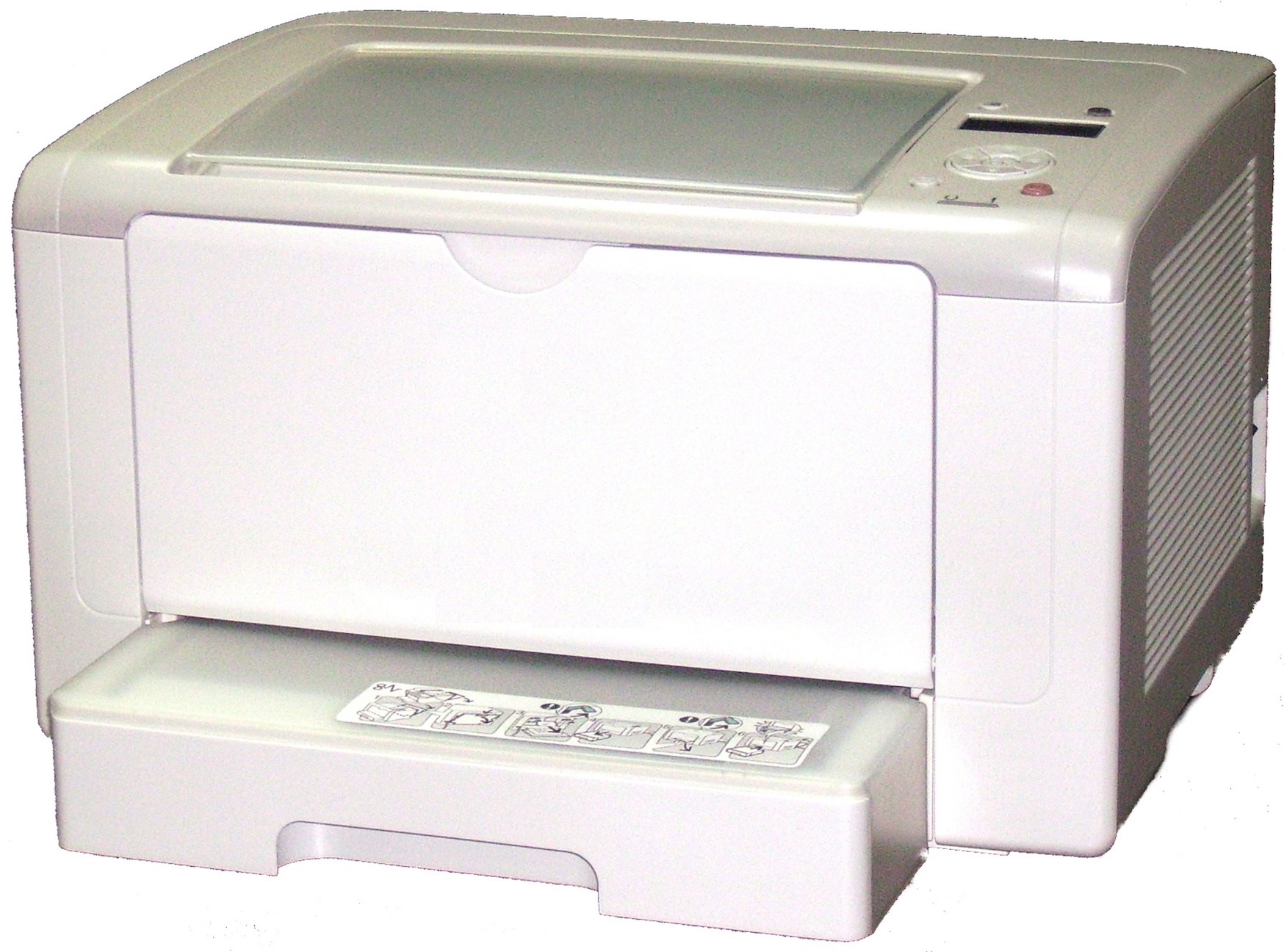}%
    \label{figure:Figure_02b}}
    \caption{Two printers, $A$ and $B$, were tested for sensitive emissions.}
    \label{figure:Figure_02}
\end{figure}

\begin{figure}[ht]
    \centering
    \includegraphics[width=\columnwidth]{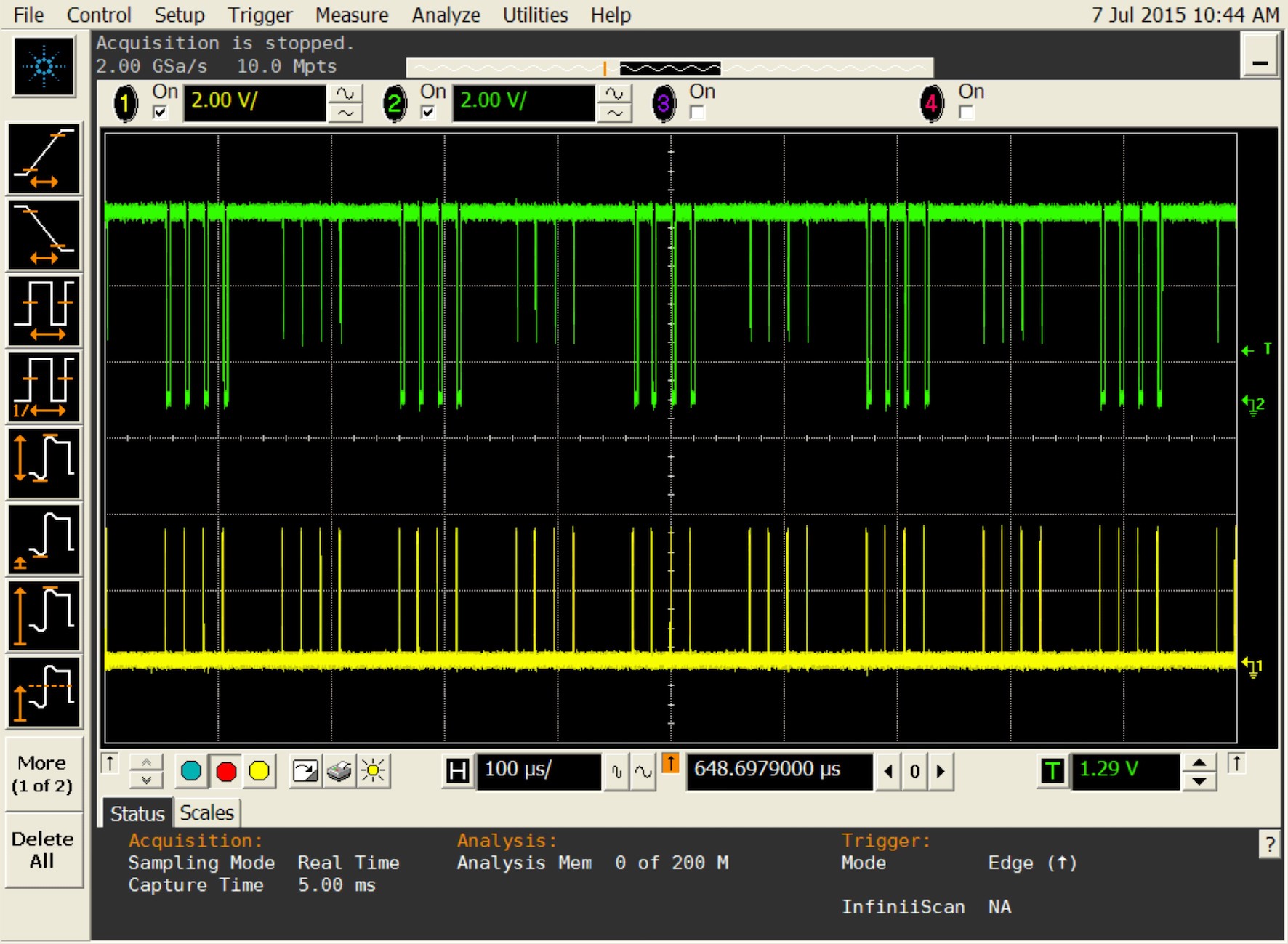}
    \caption{Waveforms of control signals on pins 6 (lower trace) and 9
    (upper trace) of printer $A$ for the 1200\,dpi mode and the Eco option.}
    \label{figure:Figure_10}
\end{figure}

For these printers, changes of operating mode print quality options do have
an effect on the structure of useful signals and thus the character of the
source of sensitive RF emissions \cite{Grzesiak2011a,Kubiak2015c}.
Information about the operating mode and print quality for printer $A$ is
encoded in the structure of the control signals (Figures
\ref{figure:Figure_07}--\ref{figure:Figure_10}). The amplitude of these
signals is approximately 4--\SI{5}{\volt}. The pulse repetition frequency
also changes depending on operating mode and the toner save option. But at
the same time, these signals carry no information about the information being
printed \cite{Kubiak2014c,Kuhn2004a}.

Moreover, the amplitude of the control signals is higher than that of the
useful signals. That could mean that control signals could be considered as a
serendipitous source of masking emissions which disturb the reception of
sensitive emanations. This phenomenon is advantageous from an electromagnetic
protection point of view \cite{Ulas2016a,Guerrieri2018a,Loughry2002a}.

\begin{figure}[ht]
    \centering
    \includegraphics[width=\columnwidth]{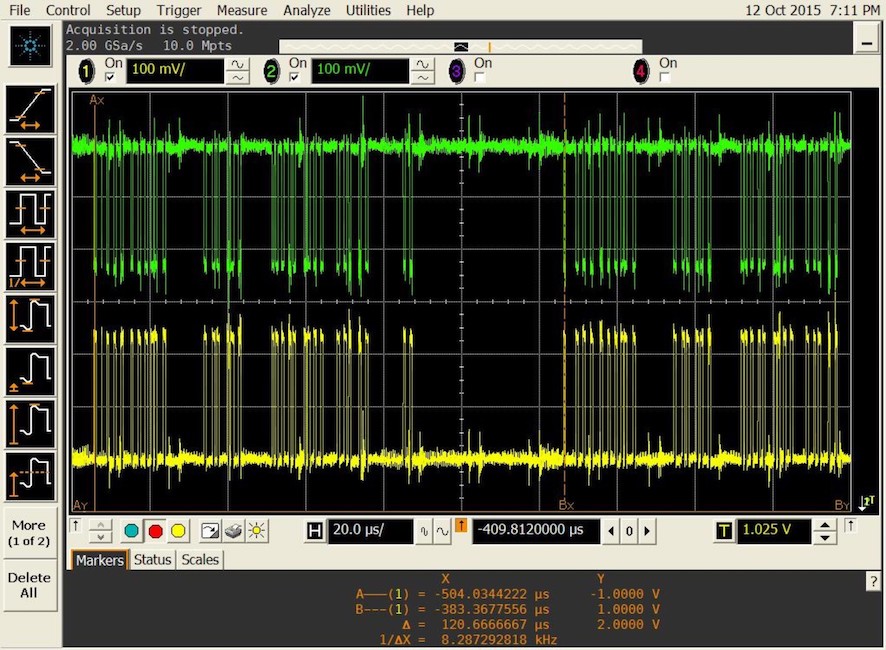}
    \caption{Waveforms of useful signals (one of the differential pairs) of
        printer $B$ for the 1200\,dpi mode and the Best option.}
    \label{figure:Figure_11}
\end{figure}

\begin{figure}[ht]
    \centering
    \includegraphics[width=\columnwidth]{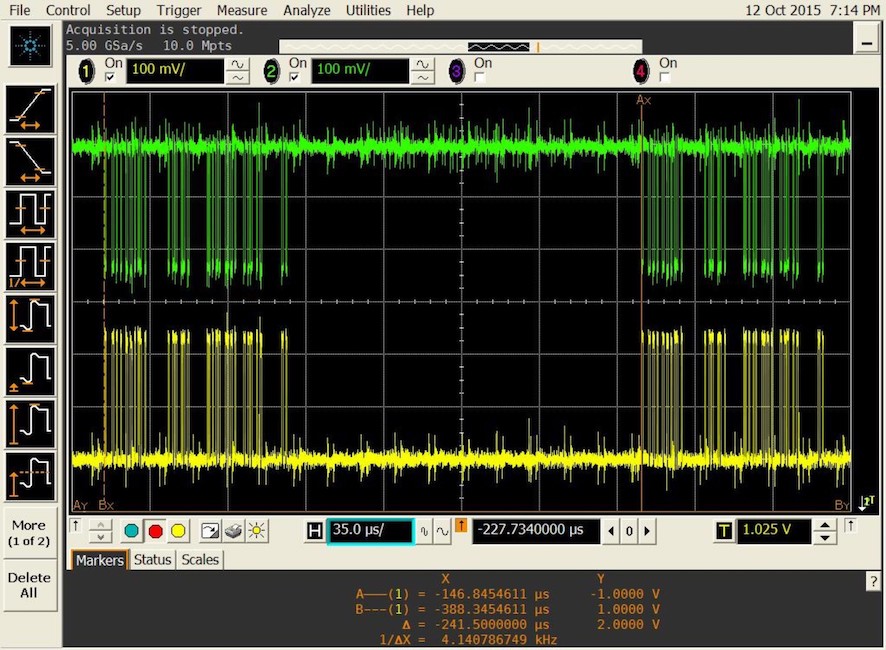}
    \caption{Waveforms of useful signals (one of the differential pairs) of
        printer $B$ for the 1200\,dpi mode and the Eco option.}
    \label{figure:Figure_12}
\end{figure}

\begin{figure}[ht]
    \centering
    \includegraphics[width=\columnwidth]{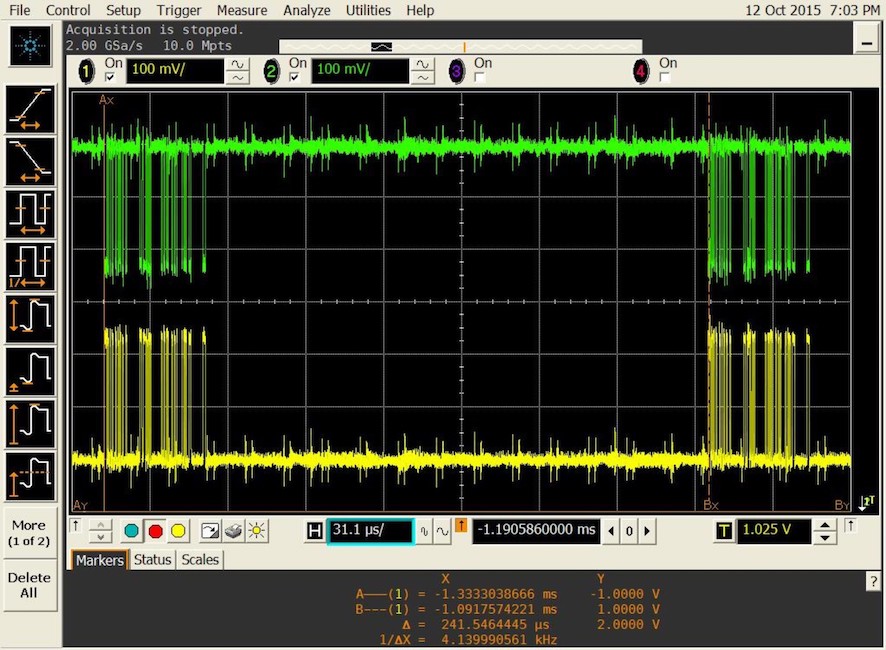}
    \caption{Waveforms of useful signals (one of the differential pairs) of
        printer $B$ for the 600\,dpi mode and the Best option.}
    \label{figure:Figure_13}
\end{figure}

\begin{figure}[ht]
    \centering
    \includegraphics[width=\columnwidth]{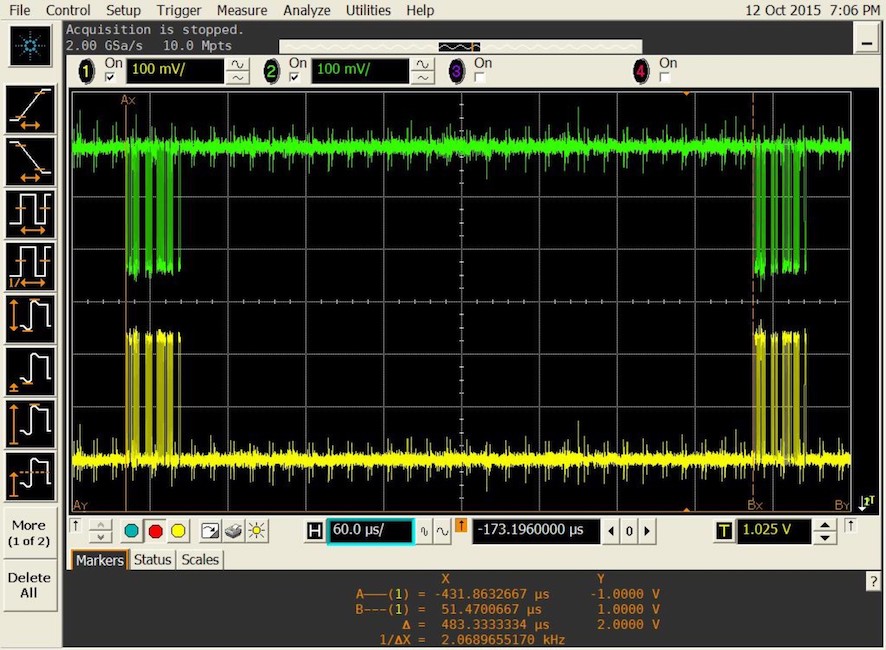}
    \caption{Waveforms of useful signals (one of the differential pairs) of
        printer $B$ for the 600\,dpi mode and the Eco option.}
    \label{figure:Figure_14}
\end{figure}

A completely different method of control of the LED array was implemented in
printer $B$ despite using the same xerographic technology of photosensitive
drum. Here, some information about modes of operation and toner save option is
visible in the useful signal. The amplitude of this signal is approximately
\SI{250}{\milli\volt}. The amplitude is less than a tenth of similar signals
in printer $A$. Moreover, the signalling method is differential.
Figures \ref{figure:Figure_11}--\ref{figure:Figure_14} show example waveforms
of useful signals. For these signals, the pulse repetition frequency changes
when printing mode of operation and printout quality are changed (Table
\ref{table:Table_2}). The structure of these signals (duty cycle) does not
change. By analysis of the parameters of useful signals we can derive an
important property crucial to reconstructing images from intercepted RF
signals that contain printed data. In the case of printer $B$, a change of
printing quality (from Eco to Best and {\it vice versa}), for a fixed
printing mode, causes predictable changes of the pulse repetition rate of the
useful signal. For printer $A$, changes to these parameters (printing mode
and printing quality) are not reflected in the behaviour of the useful
signal.

\begin{table}[H]
    \caption{Parameters of useful signals from printer $B$ in relation to
        printing parameters.}
    \label{table:Table_2}
    \centering
    \begin{tabular}{|c|c|c|c|}
        \hline
        Operating\T& PRF of differ- & First Differential & Second Differential\\
        Mode (dpi, & ential Signals & Pair [1, 3, 5, 7]  & Pair [2, 4, 6, 8] \\
        quality)\B & (\si{\kilo\hertz}) & (\si{\milli\volt}) &
                                                           (\si{\milli\volt}) \\
        \hline
        600, Eco\T   & 2.07 & $-250$ & $+250$ \\
        600, Best    & 4.14 & $-250$ & $+250$ \\
        1200, Eco    & 4.14 & $-250$ & $+250$ \\
        1200, Best\B & 8.28 & $-250$ & $+250$ \\
        \hline
    \end{tabular}
\end{table}

\section{Reconstructed Images from Sensitive Emissions}

Images of printed data were recreated from recorded (RF) useful signals
transmitted in the wires which supply signals to the LED array. The test
signal bandwidth was determined according to the equation:

$$
B = \frac{W\cdot L\cdot (\text{dpi})^2}{t}
$$

\noindent where:

\begin{description}
    \item[\textbf{\textit{B}}] is the signal bandwidth for printing one pixel,

    \item[\textbf{\textit{W}}] is the width of the printing area in inches,

    \item[\textbf{\textit{L}}] is the length of the printing area in inches,

    \item[dpi] is the printing resolution in dots per inch, and

    \item[\textbf{\textit{t}}] is the time to print one page.
\end{description}

We have to know the printing parameters to reconstruct the original
information. These parameters are: the length of printer video line (in
pixels), and the number of video lines on a sheet of paper. As we can see,
full reconstructed images can have very large dimensions; for example, for:
\begin{itemize}
    \item a resolution of $1200\times 1200$\,dpi,
    \item printing speed of 30 pages per minute,
    \item paper size of A4 (about 8.27 by 11.69 inches, that is
        $9924\times 14\,028$ pixels), and
    \item three samples per pixel collected,
\end{itemize}

\noindent we obtain a data size about \SI{450}{\mega\byte}. Therefore,
further analyses are based on fragments of images
\cite{Grzesiak2010a,Kubiak2015b}.

\begin{figure*}[ht]
    \centering
    \subfloat[]{
        \includegraphics[width=0.49\columnwidth]{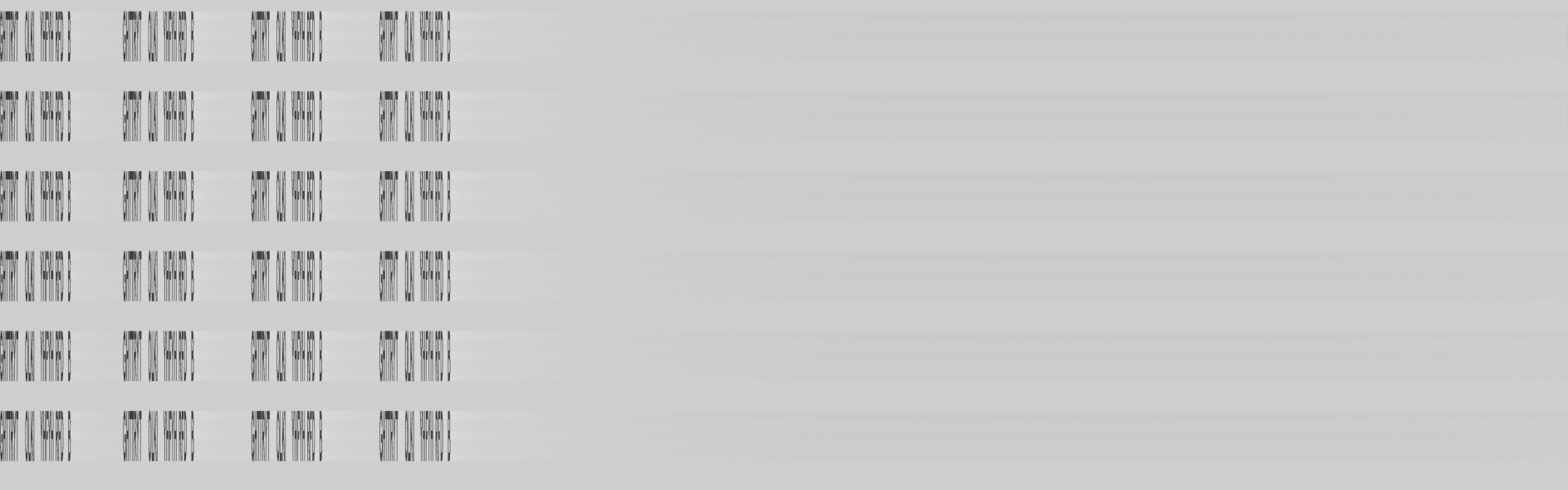}%
        \hspace{\fill}
        \includegraphics[width=0.49\columnwidth]{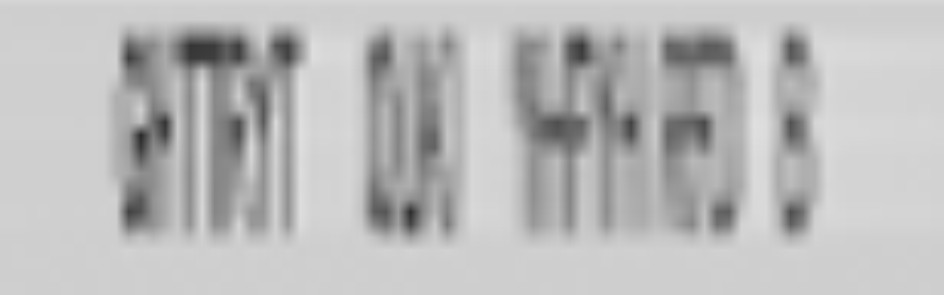}%
        \label{figure:Figure_15a}
        \hspace{\fill}
        \includegraphics[width=0.49\columnwidth]{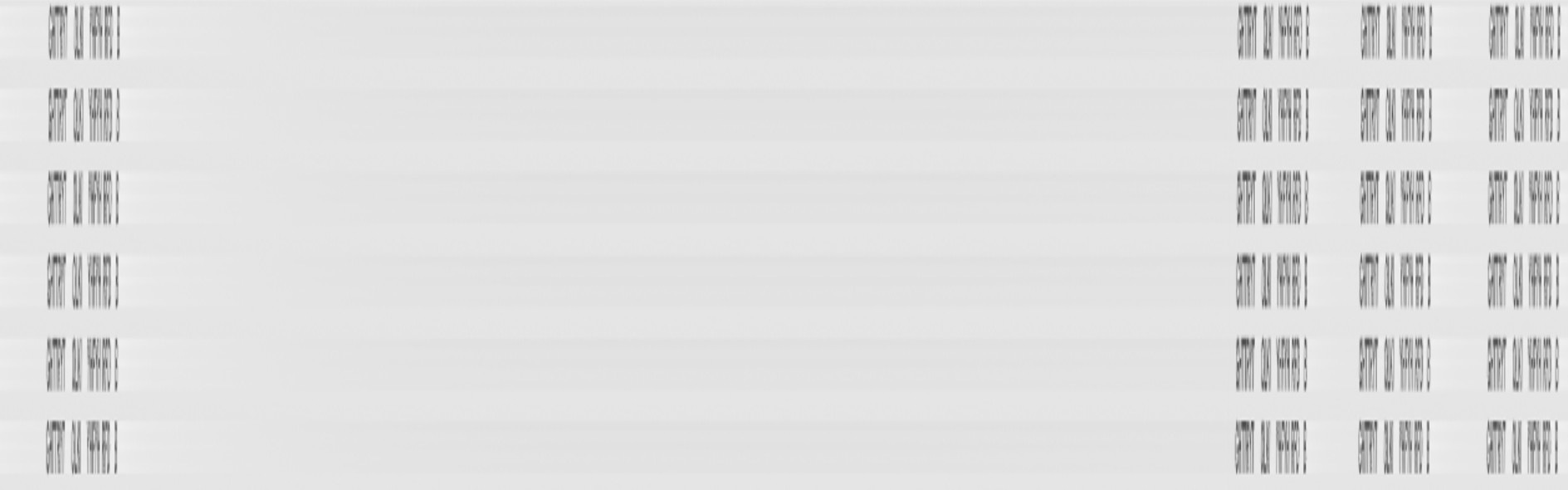}%
        \hspace{\fill}
        \includegraphics[width=0.49\columnwidth]{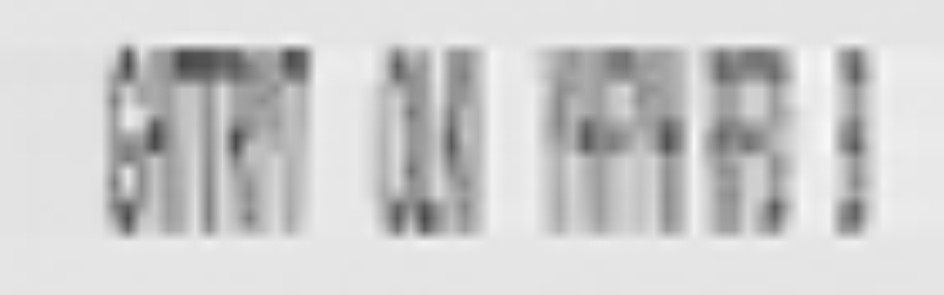}%
        \label{figure:Figure_15b}
    }
    \hfill
    \subfloat[]{
        \includegraphics[width=\columnwidth]{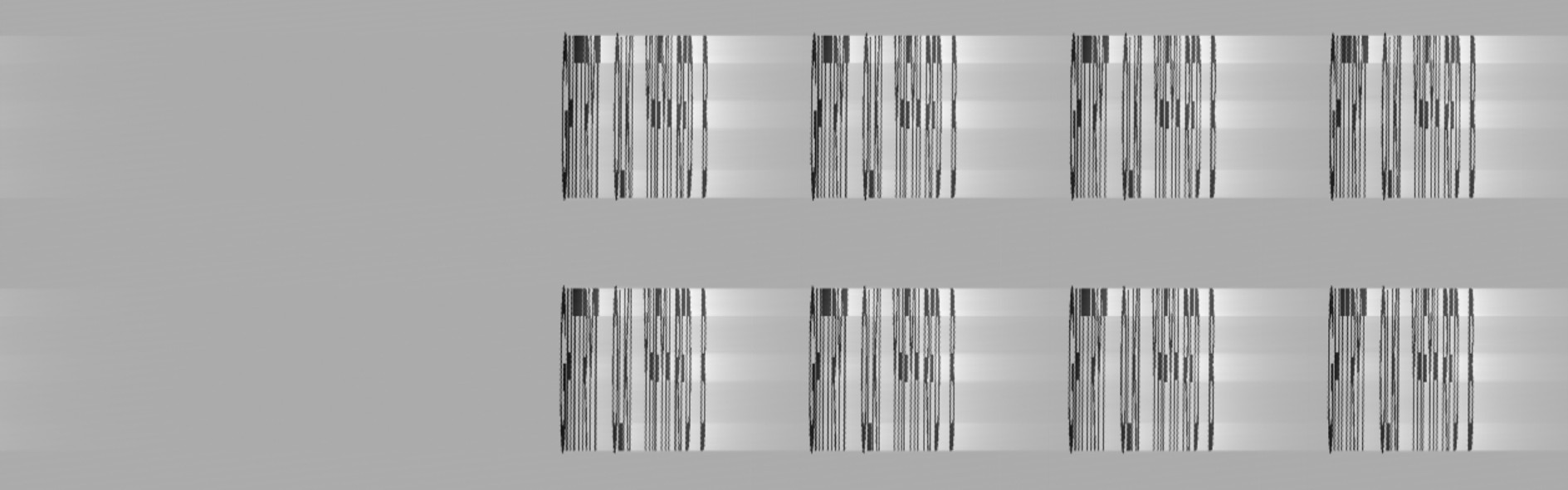}%
        \hspace{\fill}
        \includegraphics[width=\columnwidth]{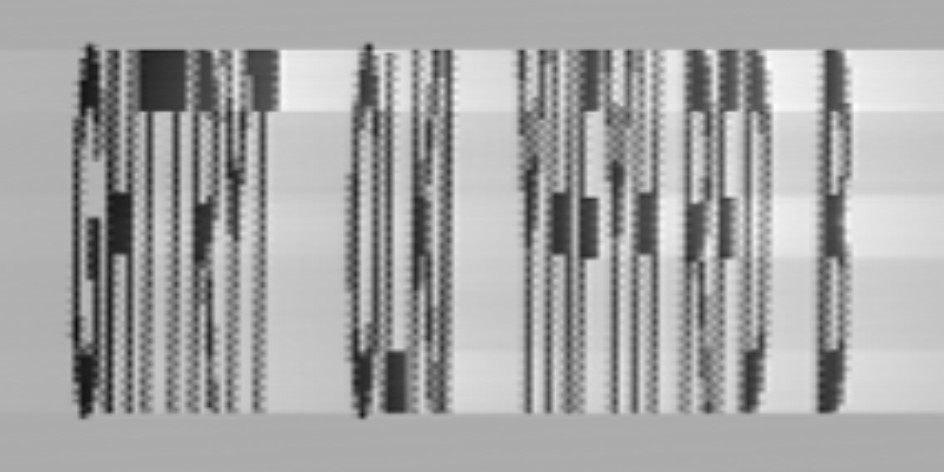}%
        \label{figure:Figure_15c}
    }

    \caption{Examples of images reconstructed from useful signals (printer
        $A$) for (a) 300\,dpi with toner save, (b) 600\,dpi without toner
        save, and (c) 1200\,dpi with toner save.}
    \label{figure:Figure_15}
\end{figure*}

\begin{figure*}[!ht]
    \centering
    \subfloat[]{
        \includegraphics[width=\columnwidth]{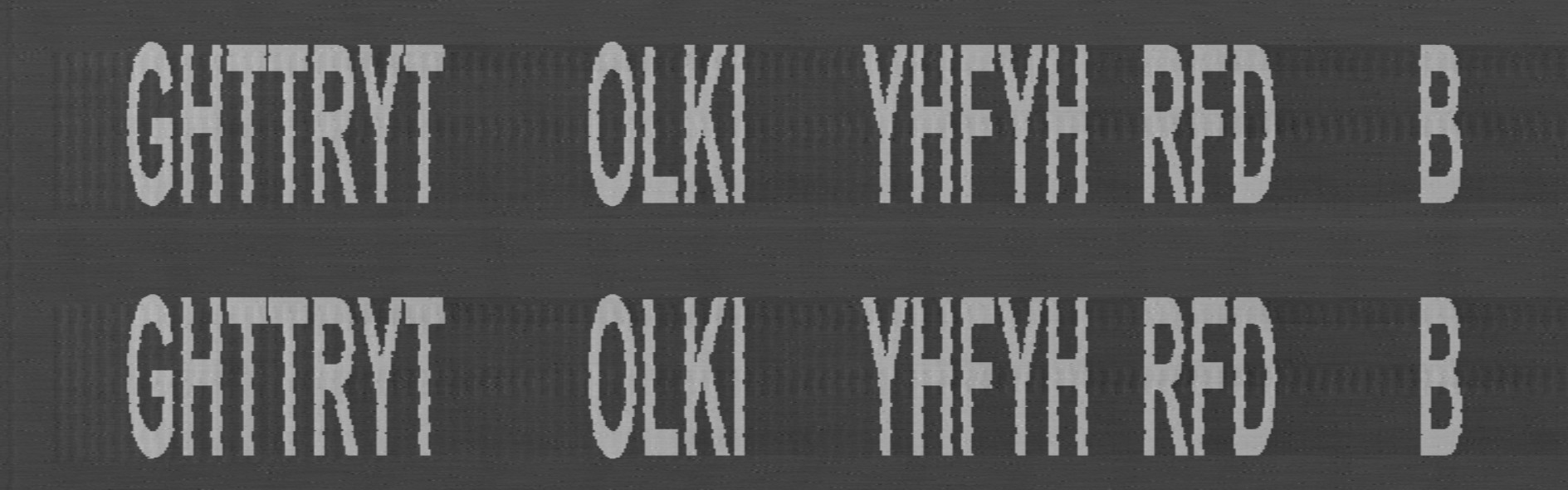}%
        \hspace{\fill}
        \includegraphics[width=\columnwidth]{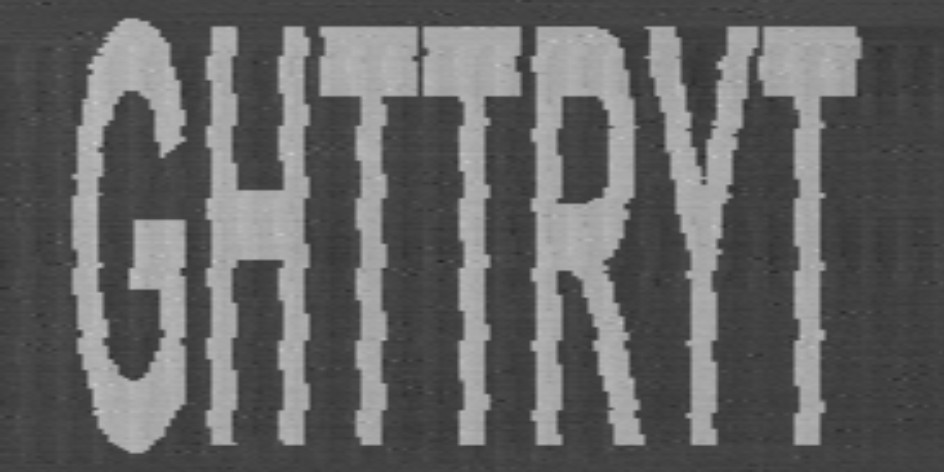}%
        \label{figure:Figure_16a}
    }
    \hfill
    \subfloat[]{
        \includegraphics[width=\columnwidth]{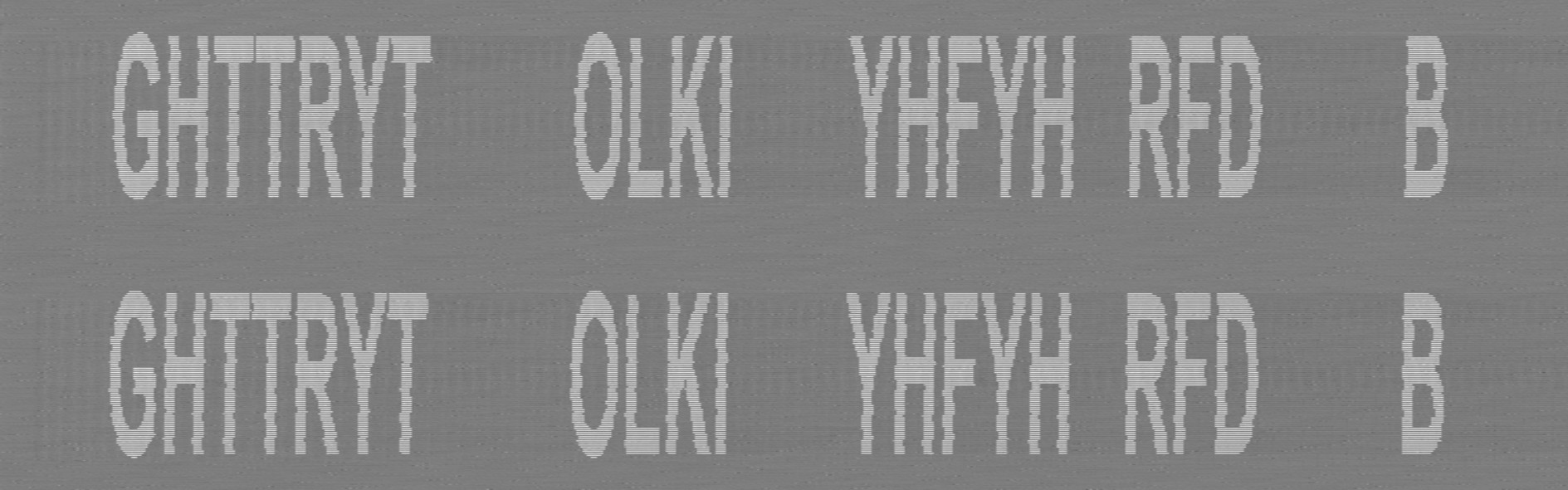}%
        \hspace{\fill}
        \includegraphics[width=\columnwidth]{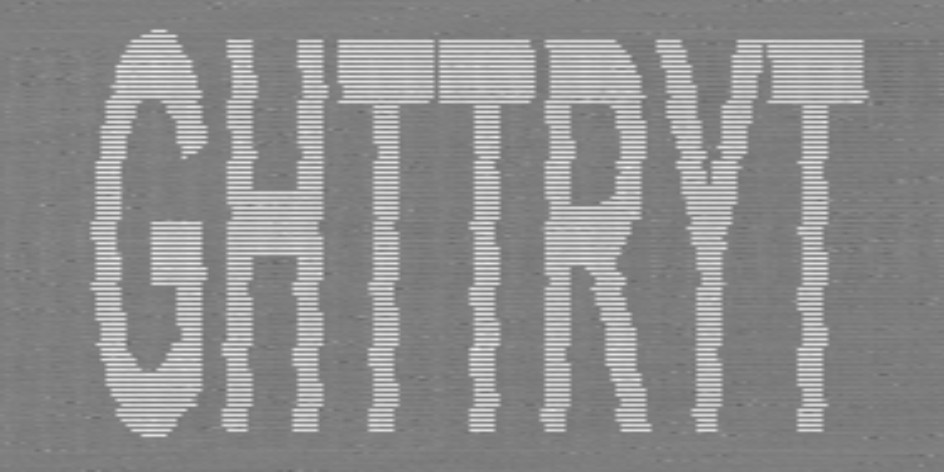}%
        \label{figure:Figure_16b}
    }
    \hfill
    \subfloat[]{
        \includegraphics[width=\columnwidth]{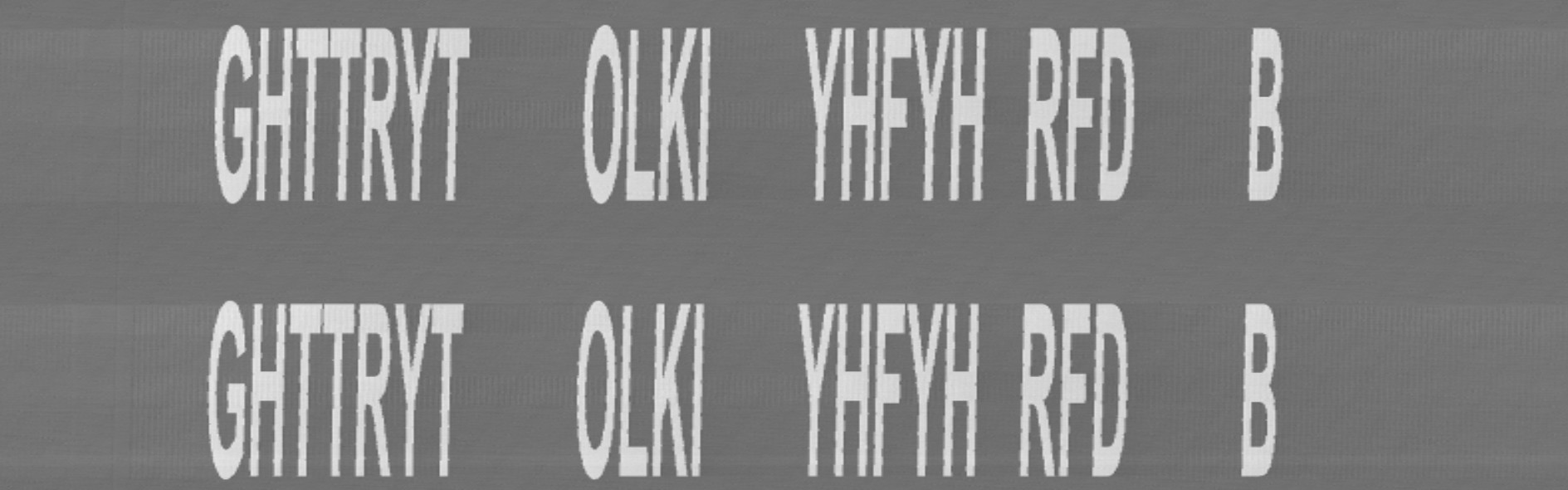}%
        \hspace{\fill}
        \includegraphics[width=\columnwidth]{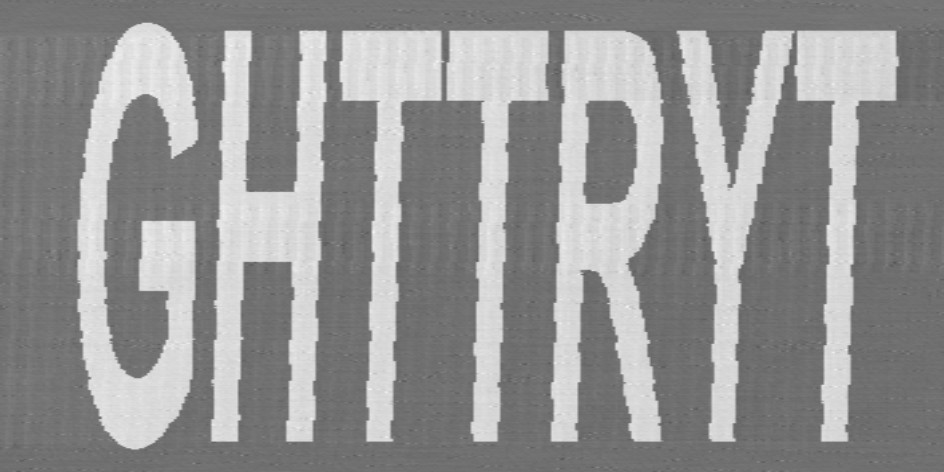}%
        \label{figure:Figure_16c}
    }
    \hfill
    \subfloat[]{
        \includegraphics[width=\columnwidth]{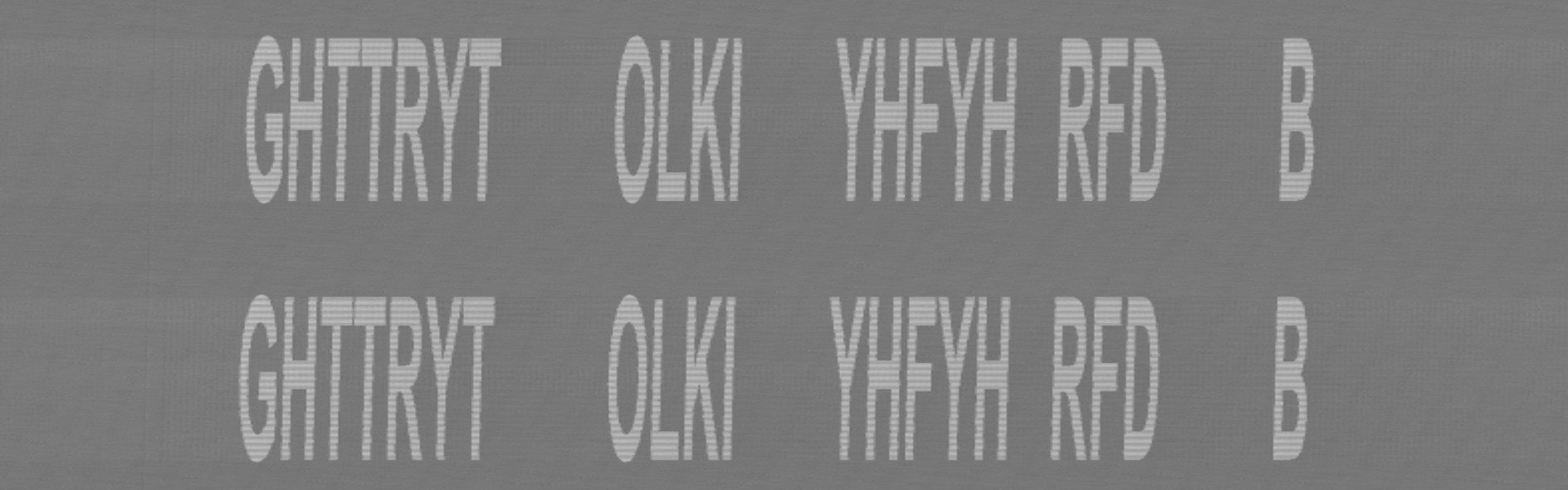}%
        \hspace{\fill}
        \includegraphics[width=\columnwidth]{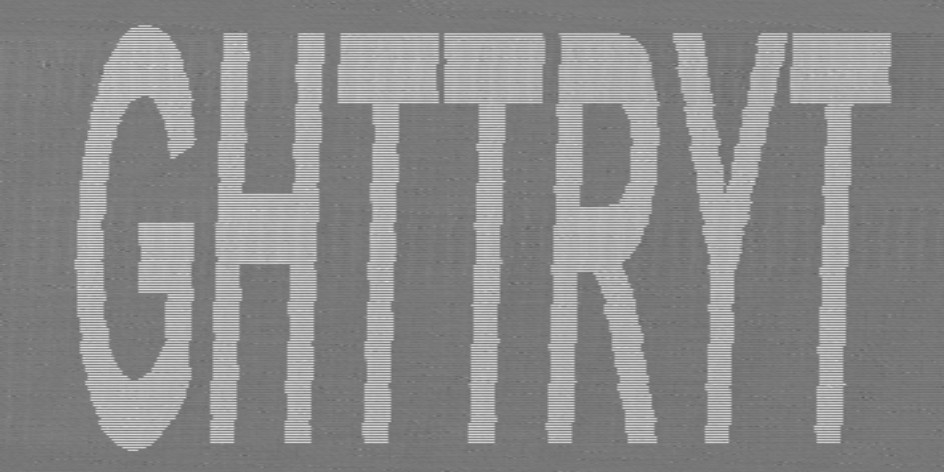}%
        \label{figure:Figure_16d}
    }

    \caption{Examples of images recreated from useful signals (printer
        $B$) for (a) 600\,dpi without toner save, (b) 600\,dpi with toner
        save, 1200\,dpi without toner save, and (d) 1200\,dpi with toner
        save.}
    \label{figure:Figure_16}
\end{figure*}

Fragments of these images are presented in Figures \ref{figure:Figure_15} and
\ref{figure:Figure_16}. The reconstructed glyphs contained in the
image are constructed from horizontal lines at intervals equal to the width
of the line \cite{Kubiak2006a} apart from repetition frequency of the useful
signal and option Best or Eco and for the default printing resolution of
printer $B$.

In two-diode laser printers, a phenomenon occurs that causes the
reconstructed images from sensitive emissions to contain only single points,
corresponding to the beginning and endpoint of each horizontal line
comprising the printed glyphs (essentially run-length encoding
the reconstructed images) \cite{Kubiak2014d}.

However, these useful signals were not differential signals. In the case of
printer $B$, despite the predictable structure of character glyphs, the
differential signalling used tends to help protect printed data against
electromagnetic infiltration \cite{Ketenci2017a}.

\section{Levels of Electromagnetic Emissions}

Printer $B$ uses differential transmission of useful signals. Its primary aim
is probably to lower the levels of electromagnetic emission and increase
resistance to external disturbances. Since the useful differential signal is
responsible for the transfer of information from the printer's raster image
processor (RIP) to the LED array, its characteristics correspond to
characteristics of the processed information. Therefore, the solution adopted
by the printer's designer (in the form of a differential signal) also reduces
the levels of electromagnetic emission correlated with printed data
\cite{Song2015a}. Such a solution was not used in printer $A$, despite the
fact that the amplitude of useful signals are over ten times higher than in
the case of printer $B$. This necessarily translates into a level of
electromagnetic emission. The number of wires carrying useful signals is half
that of printer $B$.

\begin{figure*}[ht]
    \centering
    \includegraphics[width=\textwidth]{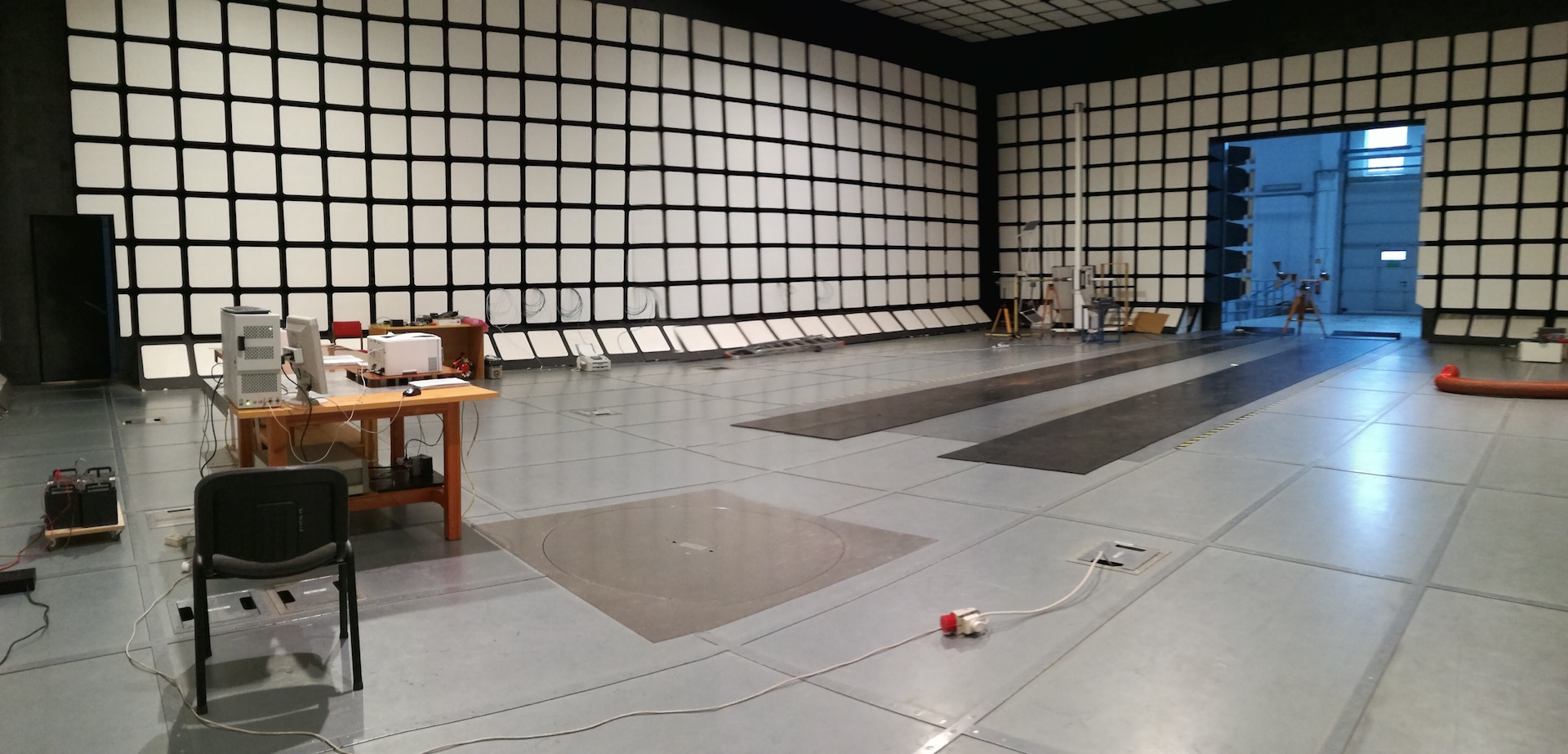}
    \caption{Anechoic chamber where the tests were carried out.}
    \label{figure:Figure_17}
\end{figure*}

\begin{figure}[ht]
    \centering
    \includegraphics[width=\columnwidth]{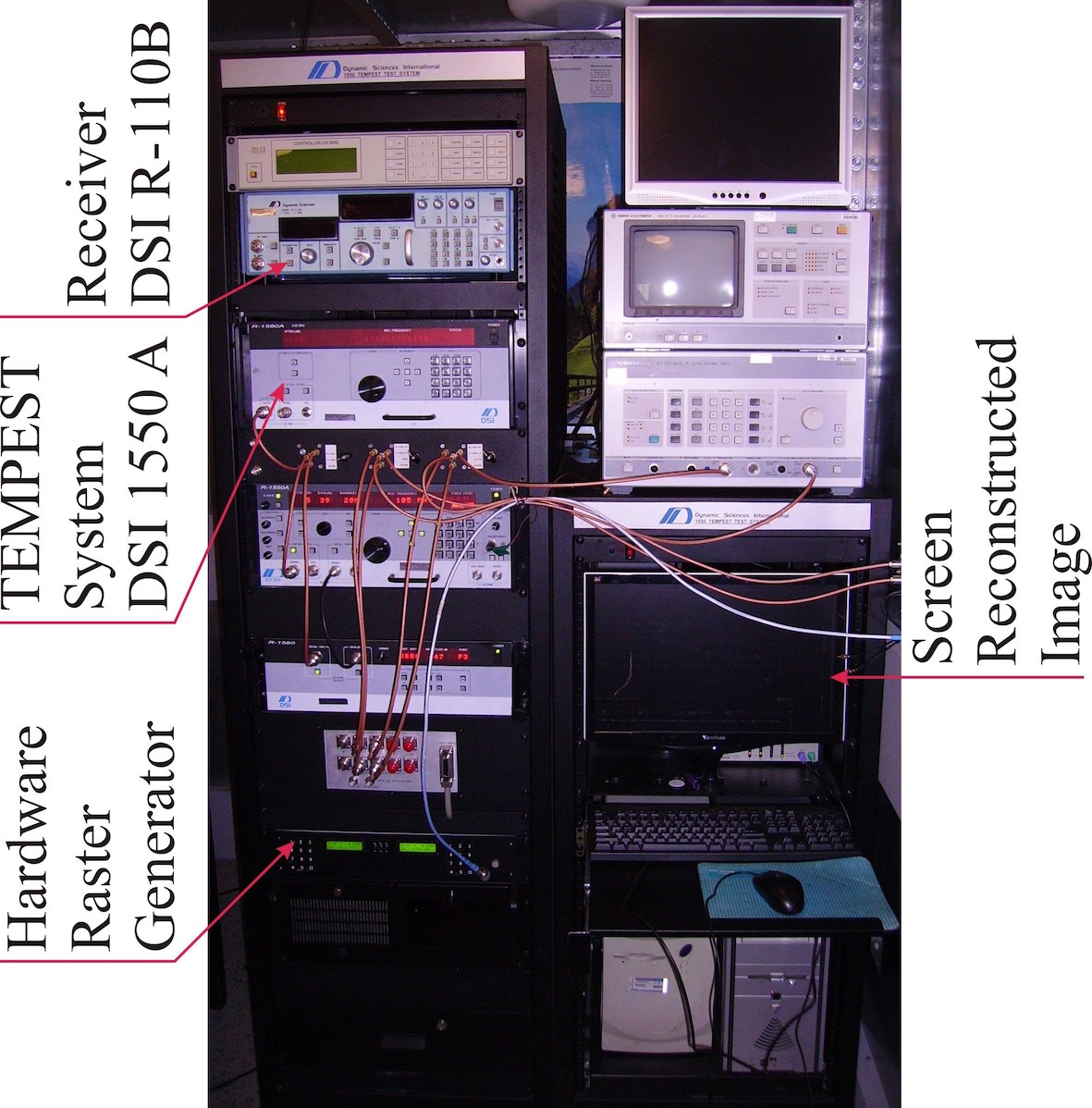}
    \caption{TEMPEST Test System model DSI 1550A.}
    \label{figure:Figure_18}
\end{figure}

An anechoic chamber (Figure \ref{figure:Figure_17}) was used to test the
validity of the assumptions.
During the tests, sensitive emissions were measured with a bandwidth of
\SI{1}{\mega\hertz} in the frequency range from \SI{2}{\mega\hertz} to
\SI{1}{\giga\hertz}. This frequency range was selected as a result of many
years of experience in testing of laser printers and display screens. The
aforementioned bandwidth value is the most effective for sensitive emissions
from laser printers. During the tests, a TEMPEST DSI 1550A receiving system
(\SI{20}{\hertz}--\SI{22}{\giga\hertz}), which can be seen in Figure
\ref{figure:Figure_18}, was used. The tested printers were connected to the
TEMPEST computer, which is certified for electromagnetic safety.

\begin{figure*}[ht]
    \centering
    \includegraphics[width=\textwidth]{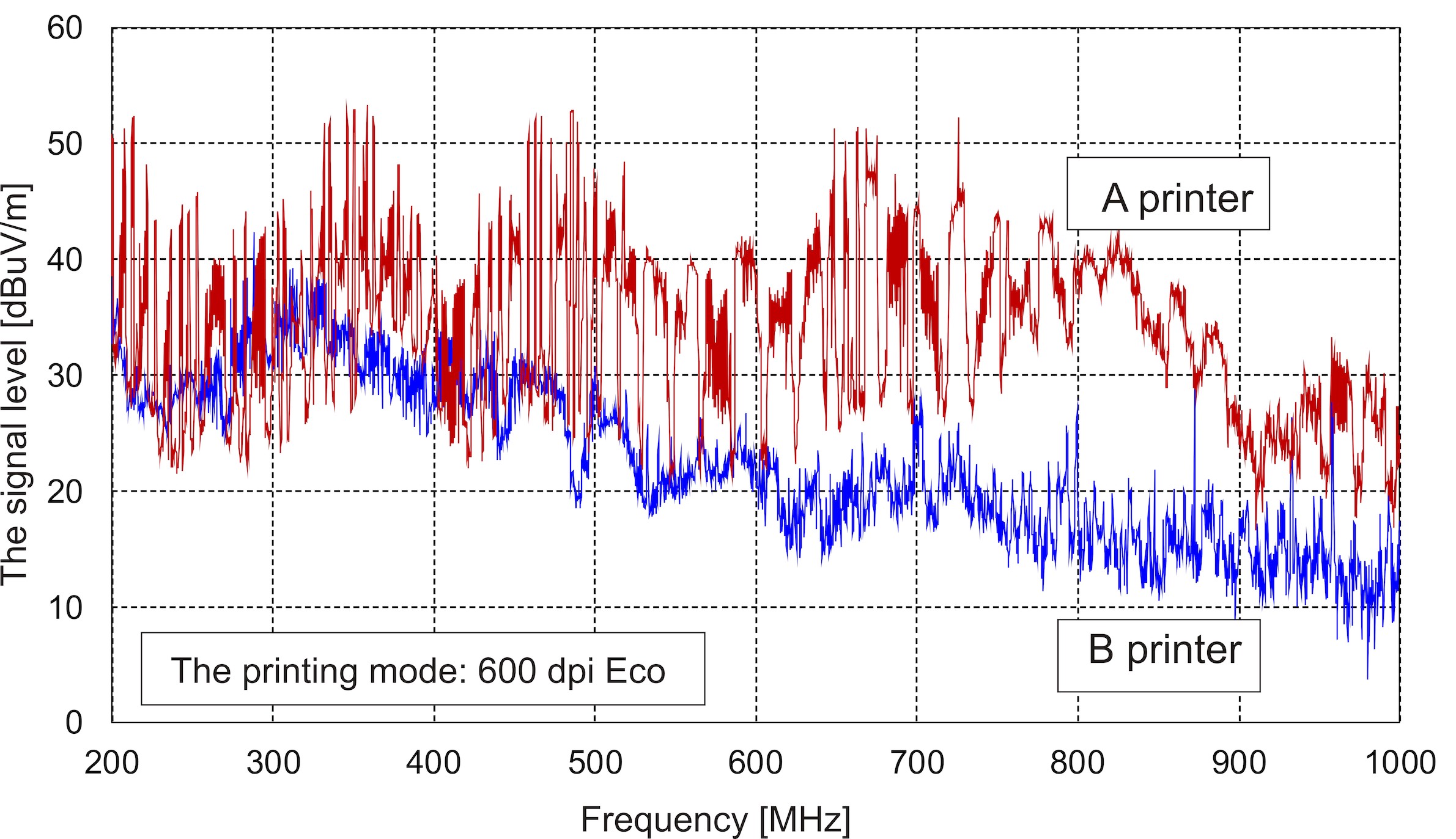}
    \caption{Radiated disturbances measured from the $A$ and $B$ printers,
        both operating in 600\,dpi mode (with Eco option); bandwidth =
        \SI{1}{\mega\hertz}.}
    \label{figure:Figure_19}
\end{figure*}

The reason for using the TEMPEST computer for this purpose is because a
typical computer has higher levels of electromagnetic emissions. These
emissions can ``cover'' the target
emissions. When that happens, the electromagnetic infiltration process
becomes impossible. Results of the TEMPEST measurements are shown in Figure
\ref{figure:Figure_19}.

\section{Sensitive Emissions}

\subsection{Printer \texorpdfstring{$A$}{A}}

\begin{figure*}[ht]
    \centering
    \subfloat[]{
        \includegraphics[width=2.38in]{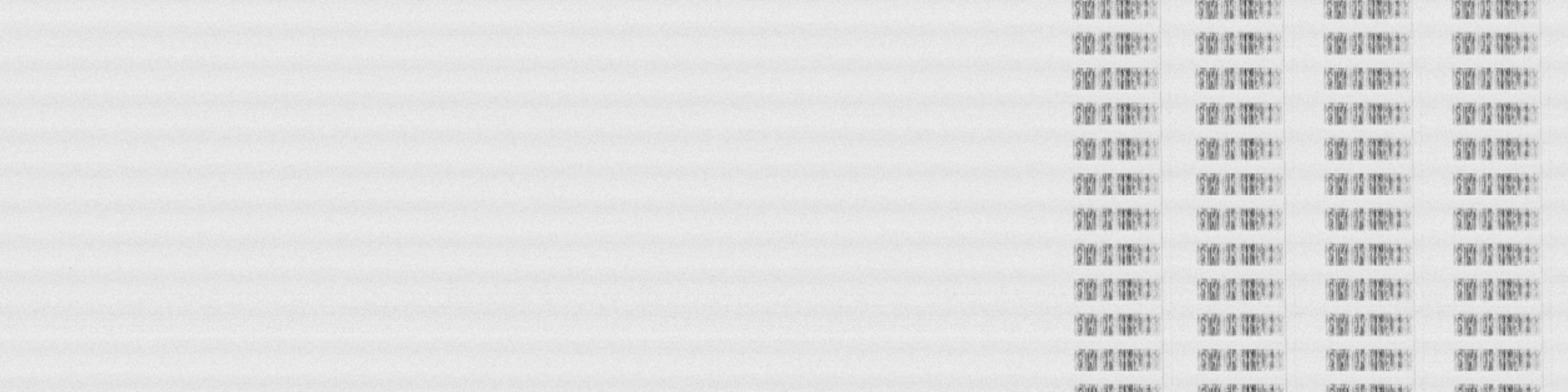}
        \label{figure:Figure_20a}
    }
    \hfil
    \subfloat[]{
        \includegraphics[width=2.38in]{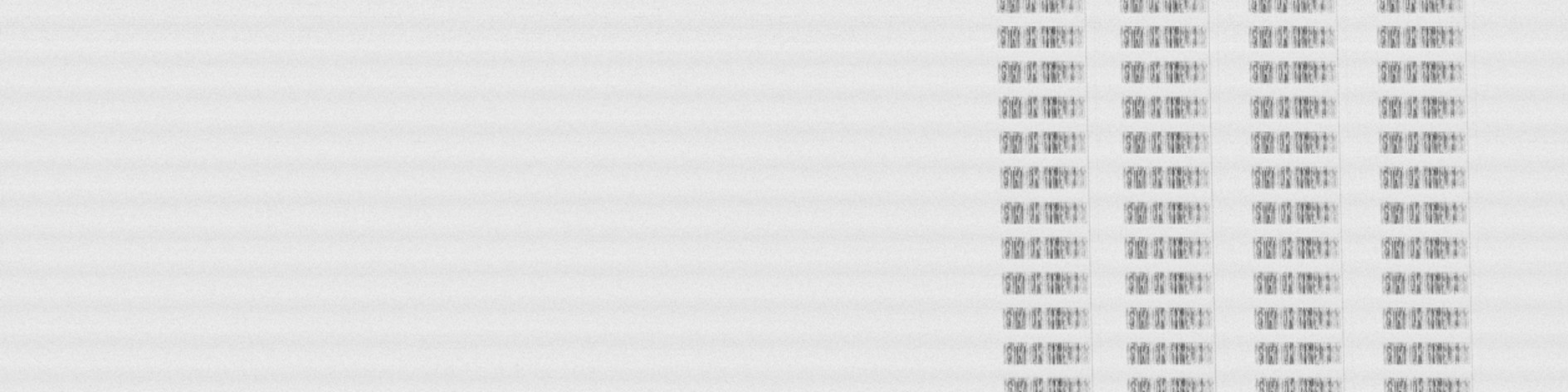}
        \label{figure:Figure_20b}
    }

    \subfloat[]{
        \includegraphics[width=2.38in]{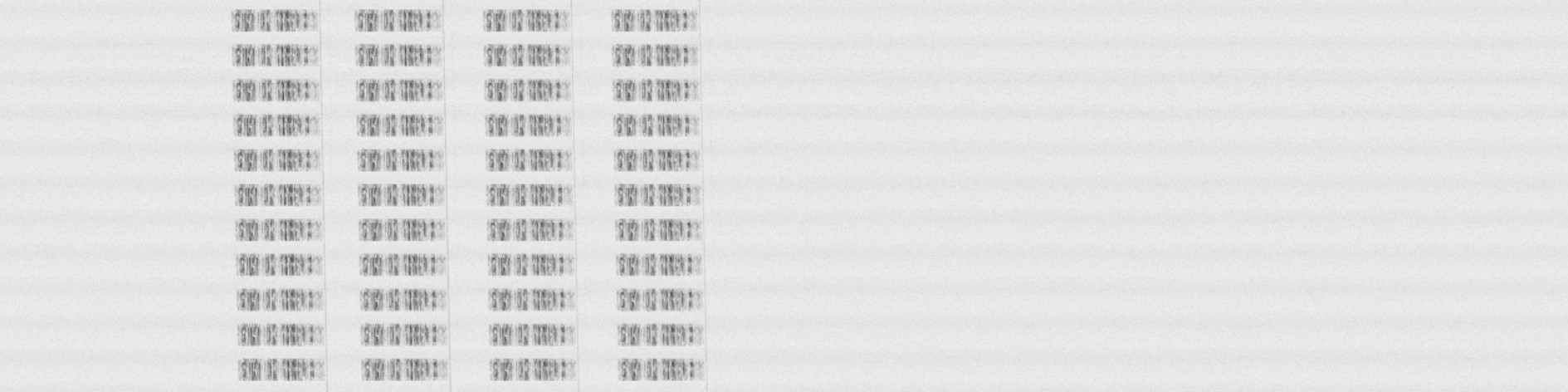}
        \label{figure:Figure_20c}
    }
    \hfil
    \subfloat[]{
        \includegraphics[width=2.38in]{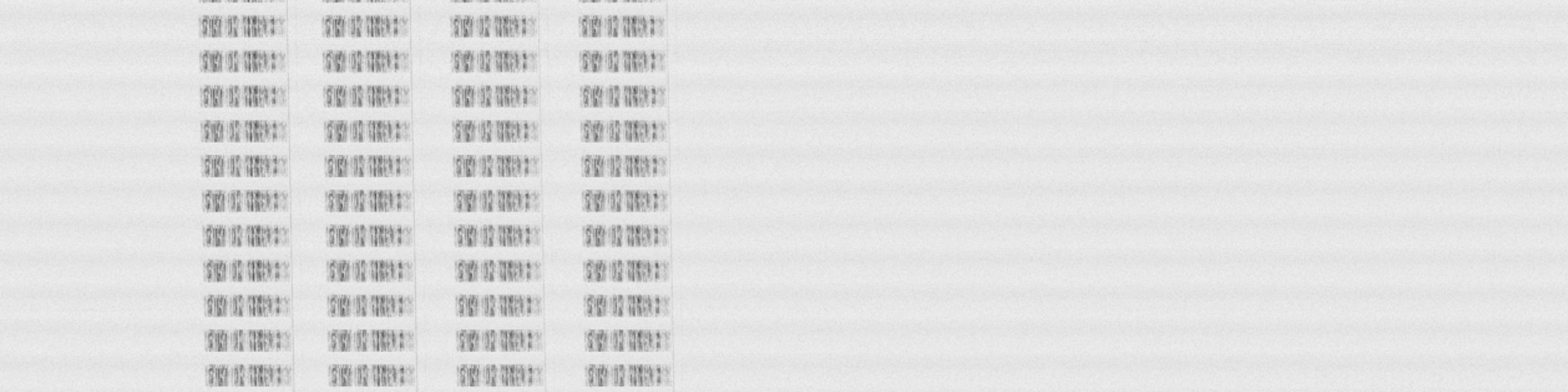}
        \label{figure:Figure_20d}
    }
    \caption{Fragments of reconstructed images from sensitive emissions of
        printer $A$: (a) 300\,dpi mode without toner save, (b) 300\,dpi mode
        with toner save, (c) 600\,dpi mode without toner save, and (d)
        600\,dpi mode with toner save. Measured frequency of sensitive
        emission: $f_0 = 525$\,\si{\mega\hertz}, BW = \SI{5}{\mega\hertz}.
        Image is inverted.}
    \label{figure:Figure_20}
\end{figure*}

\begin{figure*}[ht]
    \centering
    \subfloat[]{
        \includegraphics[width=2.38in]{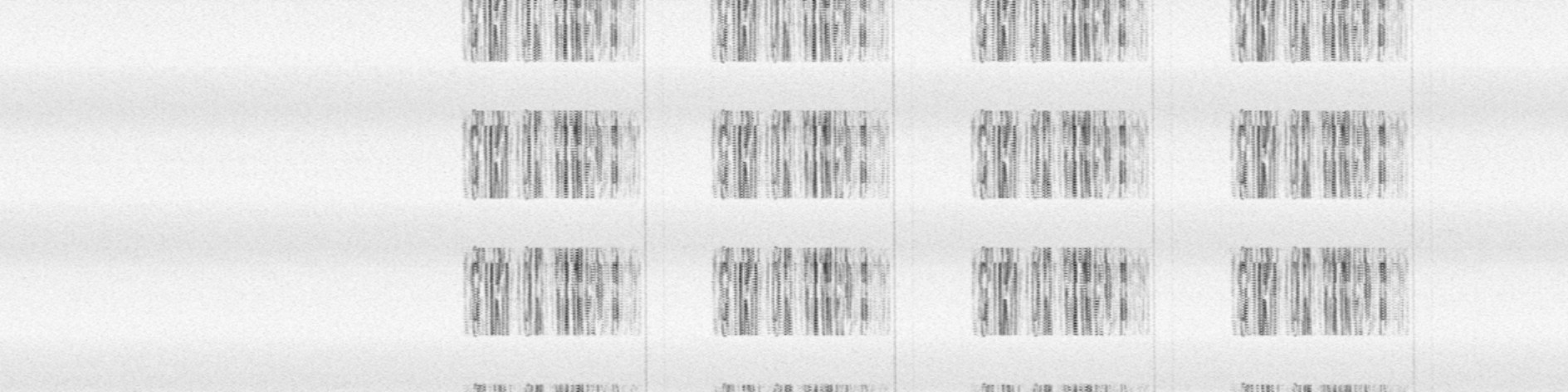}
        \label{figure:Figure_21a}
    }
    \hfil
    \subfloat[]{
        \includegraphics[width=2.38in]{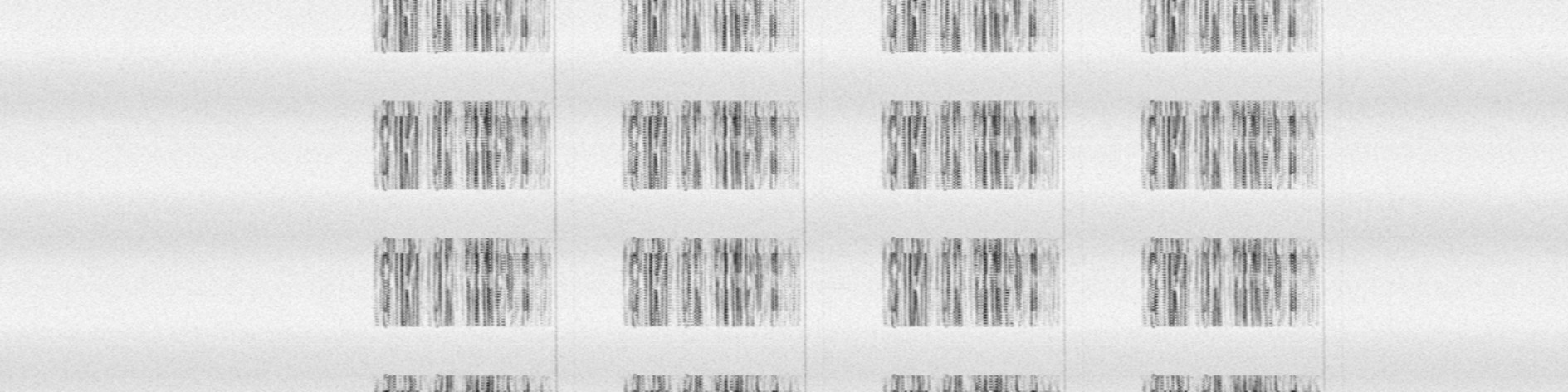}
        \label{figure:Figure_21b}
    }
    \caption{Fragments of reconstructed images from sensitive emissions of
    printer $A$: (a) 1200\,dpi mode without toner save, (b) 1200\,dpi mode
    with toner save, measured frequency of sensitive emission:
    $f_0 = 525$\,\si{\mega\hertz}, BW = \SI{5}{\mega\hertz}. Image is
    inverted.}
    \label{figure:Figure_21}
\end{figure*}

The useful signals are sent by four wires. The parameters of the signals are
constant regardless of the printing mode and toner save option. The only
change relates to the frequency of repetition of the useful signals for the
1200\,dpi mode, which is twice as high as for the two lower modes (300\,dpi
and 600\,dpi). The reconstructed images, regardless of the operating mode of
the printer, are visually similar, precluding identification of the operating
mode of the printer (Figures \ref{figure:Figure_20}--\ref{figure:Figure_21}).
At the same time, the operating mode does not change the radiated
characteristic of the emission source or the level of susceptibility to
infiltration. In this printer, information about printout quality is sent by
additional control wires. In this case, different printing modes generate
control signals having different timing structures.

As we can see, the xerographic technology of a photosensitive drum
illuminated by LED arrays, which is used in computer printers, has
significant characteristics from the point of view of electromagnetic
protection of the processed data. The reconstructed data do not directly
contain characteristics that would facilitate their identification, as is the
case with conventional laser printers for the same printing modes
\cite{Kubiak2011a}. Even the use of digital image processing---such as
extension of pixel amplitude histogram, pixel amplitude thresholding, logical
filtering, or edge detection filtering---doesn't yield satisfactory results
\cite{Kubiak2006a,Kubiak2016c,Kubiak2013a}.

\subsection{Printer \texorpdfstring{$B$}{B}}

Printer $B$ uses a similar xerographic exposure process comprising a
photosensitive drum as printer $A$. However, in the structure of the original
image, we can distinguish horizontal gaps spaced at the same interval as one
line---this is the Eco option in action---which has a positive effect on the
level of
loss of distinctive features of the original signal after passing through the
side channel attack (SCA). This printer uses differential signalling, which
feeds into the SCA a much lower amplitude original signal that is also
missing some signal features that would facilitate electromagnetic
eavesdropping.

\begin{figure*}[ht]
    \centering
    \includegraphics[width=3in]{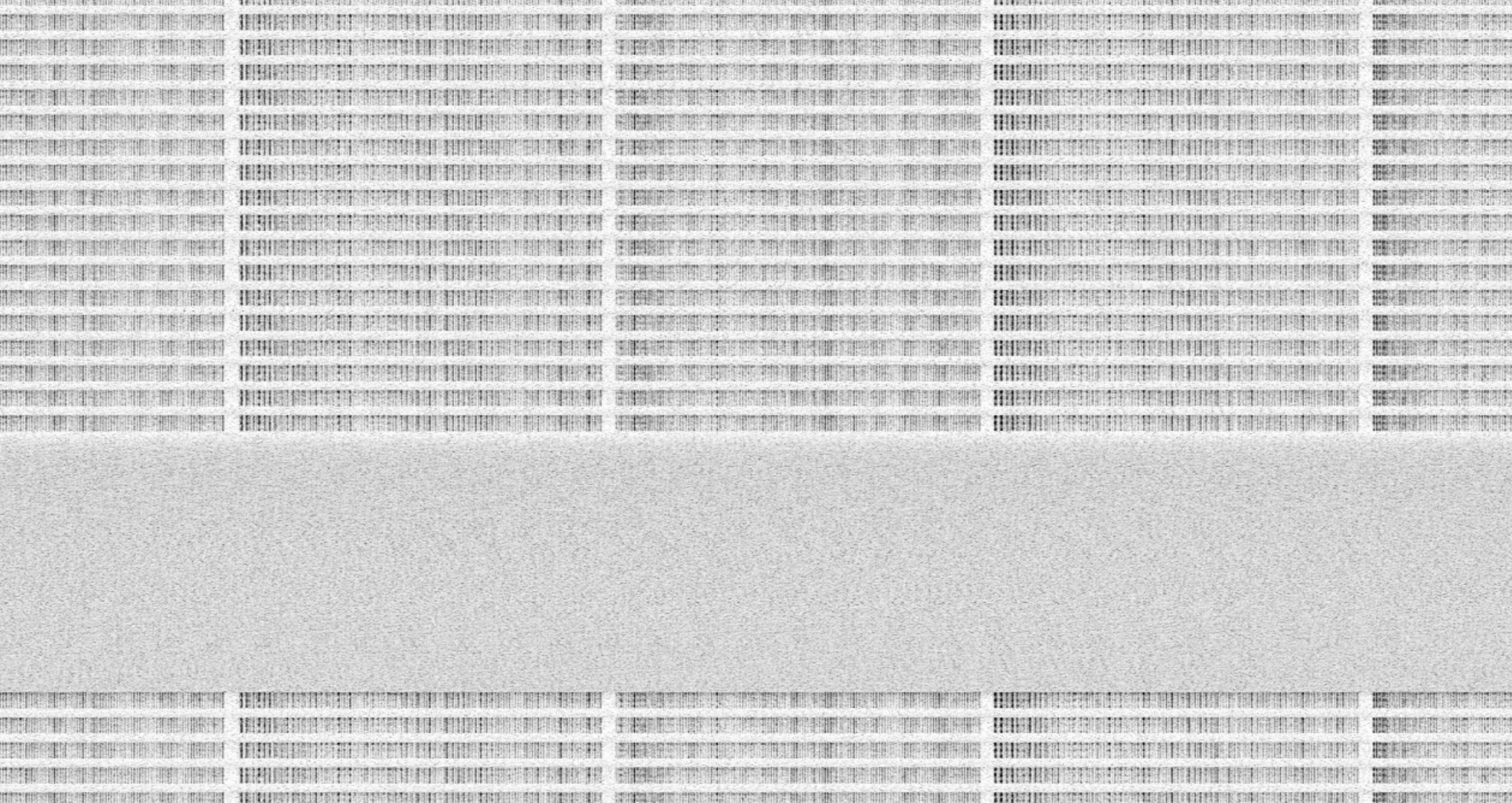}

    \bigskip

    \includegraphics[width=\textwidth]{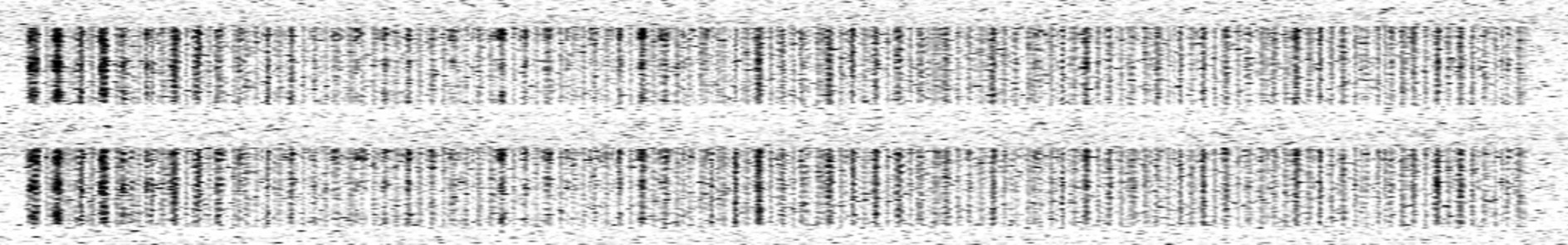}
    \caption{Printer $B$ with LED array, $600\times 600$\,dpi with toner
        save (image is inverted). Measured frequency of sensitive emission
        $f_0 = 384$\,\si{\mega\hertz}, $\text{BW}=2$\,\si{\mega\hertz}.}
    \label{figure:Figure_22}
\end{figure*}

In order to prove the conclusions above, sensitive emission signals were
recorded, and then images reconstructed from them. Undoubtedly the image
contains some glyphs \cite{Jalilian2014a}. However, due to the
elimination of a number of distinctive features caused by differential
transmission and reduction of repetition frequency of the signal (Eco
option), these elements prevent reading any information related to the
printed data (Figure \ref{figure:Figure_22}).

\section{Conclusion}

This article presents the results of tests of useful and control signals to
the LED array for the $A$ and $B$ printers. The tests were carried out from
an electromagnetic-protection-of-information point of view. The dependency of
the structures of signals on the printing mode and toner save option was
shown. In general, use of LED array technology in printers increases the
level of electromagnetic protection of information (as compared to laser
printers). The level of protection from RF electromagnetic eavesdropping is
greater than for printers employing a dual diode laser system
\cite{Kubiak2018c} and it does not require changes of construction in the
printers.

\begin{figure}[ht]
    \centering
    \includegraphics[width=3in]{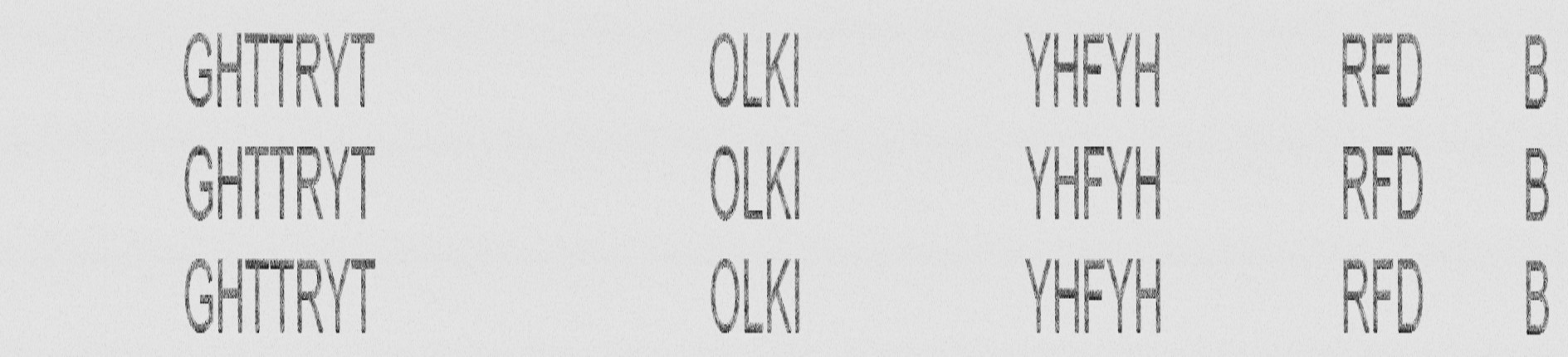}
    \caption{Example of reconstructed image from sensitive emission for a
        two-diode laser printer, 660\,dpi mode with the ``Eco'' option turned
        on. Measured frequency of sensitive emission: $f_0 =
        444$\,\si{\mega\hertz}, $\text{BW = 5}$\,\si{\mega\hertz}. Image is
        inverted.}
    \label{figure:Figure_23}
\end{figure}

Printers using the dual diode laser system use serial signal transmission.
That solution is advantageous to the electromagnetic eavesdropper (Figure
\ref{figure:Figure_23}).

\begin{table}[!ht]
    \caption[Figure 3]{Comparison of the quality of reconstructed
        data---depending on resolution (dpi) and the use of ``Best'' or
        ``Eco'' options---for laser printers that use a dual diode laser
        system or an LED array from an electromagnetic protection point of
        view.

        \medskip
        \emph{Legend:}
        \begin{description}
            \item[K1] only the edges of glyphs appear in the
                reconstructed image, but the information is legible;
            \item[K2] visible filled glyphs appear in the
                reconstructed image, but the information is not legible;
            \item[W1] visible filled glyphs appear in the
                reconstructed image, and the information is legible;
            \item[W2] glyphs in the reconstructed image are not
                visible, and the information is not legible.
        \end{description}
    }
    \label{table:Table_3}
    \centering
    \begin{tabular}{|l|c|c|c|c|}
        \hline
        & \multicolumn{2}{c|}{600\,dpi}\T\B
            & \multicolumn{2}{c|}{1200\,dpi} \\
        \hline
        \textbf{Type of Printer}\T\B & `Best' & `Eco' & `Best' & `Eco' \\
        \hline
        Dual diode printer\textsuperscript{*}\T\B & W1 & K1 & W1 & K1 \\
        \hline
        Dual diode printer\textsuperscript{\dag}\T\B & W1 & W1 & K1 & W1 \\
        \hline
        Dual diode printer\textsuperscript{\ddag}\T\B & W1 & W1 & W1 & W1 \\
        \hline
        LED array printer\T\B $A$ & K2 & K2 & K2 & K2 \\
        \hline
        LED array printer\T\B $B$ & \textbf{W2} & \textbf{W2}
            & \textbf{W2} & \textbf{W2} \\
        \hline
    \end{tabular}
    \medskip \\
    \textsuperscript{*}Producer 1 \hfill
    \textsuperscript{\dag}Producer 2 \hfill
    \textsuperscript{\ddag}Producer 3
\end{table}

The LED array system requires parallel signal transmission. This causes
successful reception and decoding of sensitive emissions to be very
difficult. The reconstructed images from valuable emissions obtained from
LED-array-based printers can be seen to contain glyphs, but they
aren't legible.

Printer $B$ goes further by using differential signalling. This method, once
adopted, significantly reduces the level of useful electromagnetic emission
(from the perspective of an eavesdropper) and thus reduces the effectiveness
of receiving emission sources. The reconstructed images cannot be read by
humans. Therefore the resistance level of printer $B$ to electromagnetic
eavesdropping is much higher than printer $A$---and that of typical
laser printers (with dual diode laser system). On the basis of recorded
signals and reconstructed images we may draw the conclusion that the method
works. However, the low quality of the recreated data stands in the way of
easy and simple interpretation.

The collected information related to printout quality and its impact on the
forms of recreated data are presented in Table \ref{table:Table_3}. The
results obtained by analysis of the $A$ and $B$ printers were compared with
results of analogous analyses of printers using a dual diode laser system. In
summary, the best approach to increase resistance to electromagnetic
infiltration is the LED array system.

\singlespacing

\bibliographystyle{IEEEtran}
\bibliography{consolidated_bibtex_file}

\vfill\noindent{\tiny Build \input{build_counter.txt}}

\end{document}

%% file: title.tex
LED Arrays of Laser Printers as Sources of Valuable Emissions for
Electromagnetic Penetration Process

%% file: abstract.tex
Protection of information against electromagnetic eavesdropping is an
important issue. Information may be derivable from the shape of an unintended
electromagnetic signal. The resulting electromagnetic emanations can be
correlated with processing of classified information. The problem extends to
computer printers. This article presents a technical analysis of LED arrays
used in monochrome computer printers and their contribution to unintentional
electromagnetic emanations. We analysed two printers from different
manufacturers, designated $A$ and $B$. The forms of useful signals and their
dependence on parameters of printing data are presented. Analyses were based
on realistic type sizes and distribution of glyphs. Pictures were
reconstructed from received radio frequency (RF) emanations. We observed
differences in
legibility of information receivable at a distance that we attribute to
different ways used by printer designers to control the LED arrays,
particularly the difference between relatively high voltage single-ended
waveforms and lower-voltage differential signals. To decode the compromising
emanations required knowledge of---or guessing---printer operating parameters
including resolution, printing speed, and paper size. The optimal RF bandwidth
for detecting individual pixels has been determined. Measurements were
carried out across differences in construction and control of the LED arrays
in tested printers, and the levels of RF emissions compared for selected
operating modes (fast, high quality, or toner saving mode) of the printing
device.

%% file: build_counter.txt
408